\documentclass[twocolumn,superscriptaddress,nofootinbib,floatfix,aps,prb,citeautoscript,longbibliography]{revtex4-2}
\usepackage[utf8]{inputenc}
\usepackage[T1]{fontenc}
\usepackage{lineno}

\usepackage{graphicx}
\usepackage{color}
\usepackage{array}
\usepackage{dcolumn}
\usepackage{bm}
\usepackage{multirow}
\usepackage{chemformula}
\let\ce\ch
\usepackage{tabularx}
\usepackage{comment}
\usepackage{tikz}
\usepackage{xstring}

\usepackage[normalem]{ulem}
\usepackage{amsmath}
\usepackage{float}
\usepackage{booktabs}
\usepackage{tabularray}
\usepackage{orcidlink}
\usepackage{nameref}
\usepackage{cleveref}
\usepackage{subcaption}

\graphicspath{{tuga_mcp_figures/}}

\newcommand{\Tc}{\ensuremath{T_\textrm{c}}}
\newcommand{\etal}{\textit{et al.}}
\newcommand{\mevat}{\,meV/atom}
\newcommand{\epsinf}{\ensuremath{\epsilon_{\infty}}}

\newcommand{\namedlabel}[2]{\expandafter\def\csname @currentlabelname\endcsname{#2}\label{#1}}

\begin{document}

\newcommand{\bochum}{Research Center Future Energy Materials and Systems of the University Alliance Ruhr and Interdisciplinary Centre for Advanced Materials Simulation, Ruhr University Bochum, Universit\"{a}tsstra{\ss}e 150, D-44801 Bochum, Germany}
\newcommand{\physics}{Faculty of Physics and Astronomy, Ruhr University Bochum, Universit\"{a}tsstra{\ss}e 150, D-44801 Bochum, Germany}
\newcommand{\coimbra}{CFisUC, Department of Physics, University of Coimbra, Rua Larga, 3004-516 Coimbra, Portugal}

\author{Paulo R. Pires\orcidlink{0009-0005-7598-4862}}
\affiliation{\bochum}
\author{Th\'{e}o Cavignac\orcidlink{0000-0002-0409-9875}}
\affiliation{\bochum}
\author{Tiago F. T. Cerqueira\orcidlink{0000-0002-4147-8129}}
\affiliation{\coimbra}
\author{Hai-Chen Wang\orcidlink{0000-0002-2892-5879}}
\affiliation{\bochum}
\author{Silvana Botti\orcidlink{0000-0002-4920-2370}}
\email{silvana.botti@rub.de}
\affiliation{\bochum}
\affiliation{\physics}
\author{Miguel A. L. Marques\orcidlink{0000-0003-0170-8222}}
\email{miguel.marques@rub.de}
\affiliation{\bochum}
\date{\today}

%\title{A perspective on the future of materials science: Agentic materials discovery}
\title{Materials research with a general-purpose agent}

\begin{abstract}
{\centering\textbf{Abstract}\par}\vspace{0.5\baselineskip}

Over the last decade we have built Alexandria, an open database of about 15 million
density-functional calculations of crystals. It includes PBE, PBEsol, SCAN and TB-mBJ results,
structures relaxed at pressures up to 150\,GPa, dielectric tensors and about 145,000
electron--phonon calculations, together with more than 400 million structures relaxed with
machine-learned potentials. On these data we trained a generative model for crystal structures, a
universal interatomic potential and graph neural networks for more than ten properties, from
band gaps to superconducting critical temperatures. From the convex hull we computed a reaction
graph of the 17.4 billion reactions between its 193,416 stable compounds. Here we give all of this
to a general-purpose language-model agent. Working unattended, the agent screens the database for a
solar absorber, drops a candidate whose synthesis releases too little energy, and writes a
synthesis plan with precursors, crucible and atmosphere. Asked about the well-studied Ba--Zr--S
system, it finds an unreported sulfide and designs the experiment that would confirm or rule it
out. Asked for an air-stable alternative, it designs one and tests it against new structures from
the generative model. With web search alone, the same agent proposes what the literature already
discusses. Put together, the data, the models and the reaction graph let a general agent do what
none of them does alone: go from a question in plain words to an experiment that a laboratory can
run and test.
\end{abstract}

\maketitle
\twocolumngrid

%=======================================================================
\section{Introduction}
\label{sec:intro}
%=======================================================================

Over the last fifteen years materials science has become, in part, a data science.
High-throughput density-functional theory (DFT) has filled databases with millions of computed
compounds~\cite{Jain2013,Saal2013,Curtarolo2012,Schmidt2022,Schmidt2023,Cavignac2025}.
Machine-learning models trained on these data predict properties in a fraction of a
second~\cite{Schmidt2019,Pires2026}, generative models propose crystal structures that nobody has
written down before~\cite{Merchant2023,Zeni2025,Antunes2024,DeBreuck2025,DeBreuck2026}, and
universal interatomic potentials now relax structures and compute phonons close to DFT
accuracy~\cite{Kaplan2025,Loew2025,DeBreuck2026b}. A list of stable compounds with a given band gap takes one
database query.

This has moved the slow step. Deciding which of the listed compounds are real, which calculations
to distrust and how a compound could be made needs knowledge that sits in old papers and in the
habits of experimental groups. At the same time, language models have learned to use tools. In an agent loop the model writes a
plan, calls a program, reads the result and decides on the next step~\cite{Yao2023}. Agents of
this kind have planned and run chemical reactions~\cite{Boiko2023,Bran2024}, answered questions
from materials databases~\cite{Chiang2024}, written research papers~\cite{Lu2024}, proposed
hypotheses that were then tested in the laboratory~\cite{Gottweis2025}, and analysed data sets for
hours while reading more than a thousand papers per run~\cite{Mitchener2025}. Several reviews
follow the field~\cite{Jablonka2023,Ramos2025,Zhang2026,Veiga2026}, and on chemistry questions the
best models now do better on average than human experts~\cite{Mirza2025}.

In 2026 multi-agent systems wrote the first complete computer-checked proof of Fermat's Last
Theorem~\cite{Anthropic2026} and reported that the three-dimensional Navier--Stokes equations with a
smooth external force can develop a singularity in finite time~\cite{OpenAI2026,Kakaes2026}. In
both cases each step could be checked, by a proof assistant or by a referee.

The situation in materials science is more complicated. There is no proof
assistant here: to verify that a proposed compound is stable one needs at least a DFT calculation,
and to verify a synthesis one needs a laboratory and often weeks of work. Theory and
experiment also do not always agree. Formation energies from standard exchange-correlation
functionals have typical errors of tens of meV/atom~\cite{Bartel2022}, which is of the same order
as the stability margin of most compounds, and many compounds synthesized in the laboratory lie
above the computed convex hull~\cite{Sun2016}. An agent that relies only on its training
data and on web searches can do little more than repeat what is already in the literature, and it
has no means to check what it proposes.

Several groups have therefore given agents the means to run simulations, from DFT
workflows~\cite{Kumar2026,Wang2025,YangP2026} and atomistic
simulations~\cite{Ghafarollahi2025,Ghafarollahi2026} to generative and property
models~\cite{Takahara2025,Nduma2025} and curated workflows for coding agents~\cite{Deng2026}. The
Model Context Protocol (MCP)~\cite{MCP2024} has become a common way to offer such tools to an
agent~\cite{Pan2025,MPMCP2025}. However, a frontier coding agent with a shell and a cluster can
already write, run and debug a DFT calculation by itself. What it cannot reach is the data that are
not on the web: the curated results of years of calculations, the models trained on them, and a
way to move through them that a chemist would recognise.

One example is synthesis. Text-mined recipes~\cite{Kononova2019} have been used to recommend
precursors by similarity to known targets~\cite{He2023}, and language models, fine-tuned or not,
now predict precursors and firing temperatures from the
literature~\cite{Kim2024,Prein2025,Song2025}, in one case in a closed loop with
experiment~\cite{Jeon2026}. These approaches learn from what was done before. A complementary
approach starts from thermodynamics. Reaction networks built on computed energies enumerate the
reactions in a chemical system and rank them by driving force and
selectivity~\cite{Aykol2019,McDermott2021,Aykol2021,McDermott2023}. Rules read off the phase
diagram select precursors that avoid stable intermediates~\cite{Chen2024}, and active learning
updates the choice of precursors after each failed experiment~\cite{Szymanski2023b}. These
networks are usually built for one chemical system at a time.

In this work we give a general-purpose agent everything our group has built over the last decade:
the Alexandria database~\cite{Schmidt2022,Schmidt2023,Cavignac2025}, the generative model
Matra-Genoa~\cite{DeBreuck2026}, the universal interatomic potential Tuga-FF~\cite{DeBreuck2026b},
the property models of Tuga-SP~\cite{Pires2026}, and a reaction graph with every reaction between
the stable compounds of the Alexandria convex hull. We connected all of this, together with tools
for the literature and for laboratory facts, to Claude Code through MCP. We did not write a new
agent: the model, its loop and its reasoning are those of a commercial coding assistant. What
changes is what the agent can reach, and with all these resources in one place it goes in a few
minutes from a question in plain words to a candidate compound and a synthesis plan. We first
describe these resources, with the reaction graph, the newest of them, in some detail, and then
show three examples in which the agent works unattended.

%=======================================================================
\section{Results}
\label{sec:results}
%=======================================================================

\begin{figure*}[t]
    \centering
    \includegraphics[width=\textwidth]{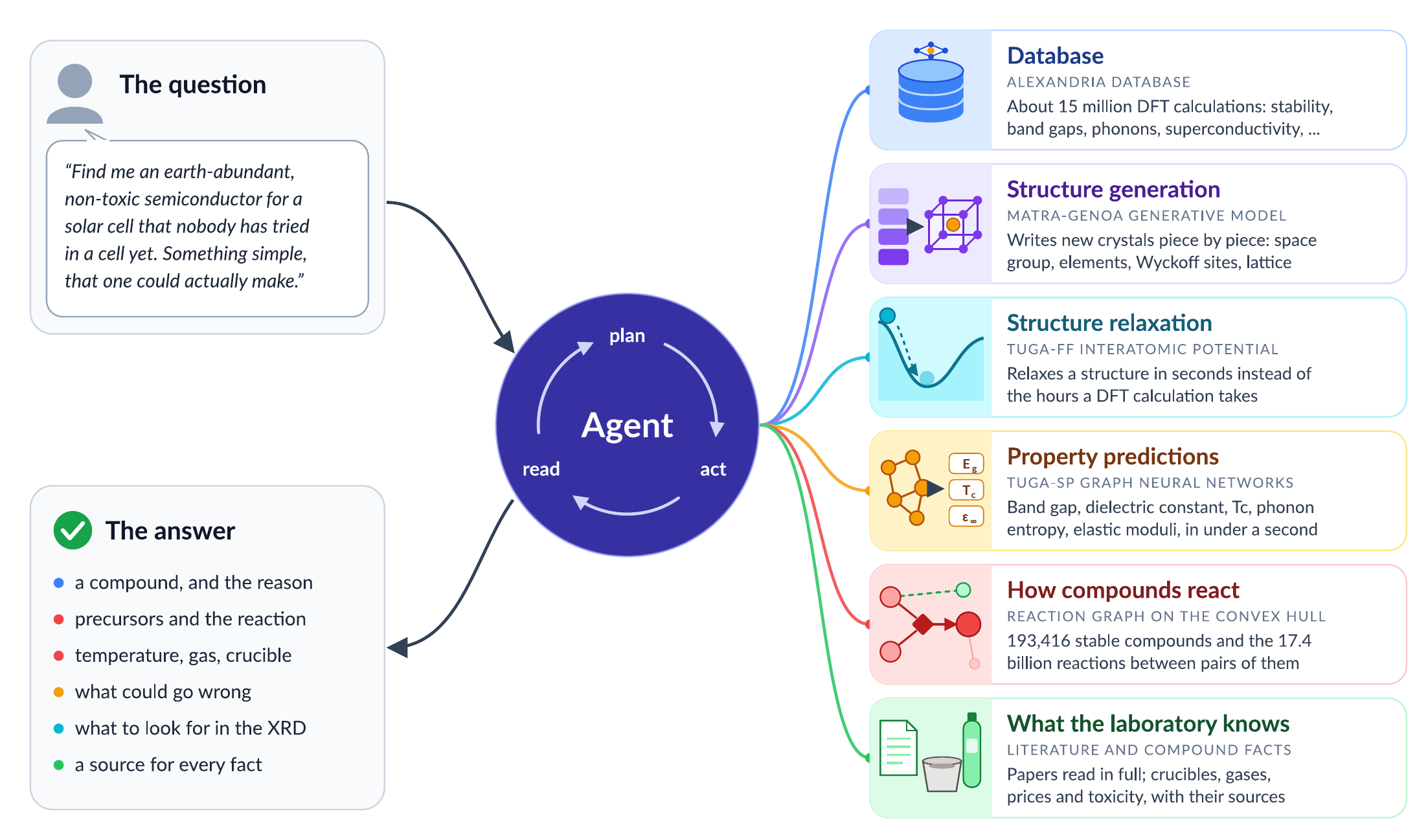}
    \caption{Schematic of the agent and its tools. The agent, a language model running in
    a loop, plans, calls one of the tools on the right, reads the answer and decides what to do
    next, until it can answer. The six resources are the data and models of our group, each
    behind a short description written for the model. The question is the one of the first
    example (Fig.~\ref{fig:example}).}
    \label{fig:concept}
\end{figure*}

\subsection{The agent and its tools}

Our agent follows the same arrangement of tools as other agents for materials science~\cite{Bran2024,Nduma2025,Deng2026} (Fig.~\ref{fig:concept}).
A researcher asks a question in plain words. The agent decides which tools to call, reads what they return and continues until it can answer. Each tool is a small program with a description written for the model about what it does, what its numbers mean and where they fail. Our agent has six kinds of resource: the Alexandria database, a generator of crystal structures, an interatomic potential, property models, the reaction graph, and the literature together with the written knowledge of our group.

With these tools an agent can work on materials discovery by itself. What we want to show is
that, more than the tools themselves, it is access to the data that makes the agent perform well.
The same tools can be used from any agent harness that speaks MCP. Each tool should also say what
it can and cannot do and, more importantly, how large the errors of the underlying theory and the
biases of the data are. A tool whose description does not say when its numbers are wrong will
be used wrongly, and the agent will repeat the mistake with confidence.

\begin{table}[t]
\caption{Content of the Alexandria database available to the agent, as numbers of entries on
28 October 2026. MatPES and MP denote the MatPES and Materials Project (GGA+$U$) settings; the
phonons and elastic constants were computed with a machine-learned interatomic potential (MLIP).}
\label{tab:alexandria}
\begin{ruledtabular}
\begin{tabular}{@{}llr@{}}
Data & Method & Entries \\
\colrule
3D crystals & PBE (MatPES) & 3.8\,M \\
3D crystals & PBE (MP) & 7.1\,M \\
3D crystals & PBEsol; SCAN & 441\,k; 434\,k \\
3D, 25--150\,GPa & PBE (MatPES) & 2.7\,M \\
2D; 1D materials & PBE (MatPES) & 159\,k; 13\,k \\
Band gaps & TB-mBJ & 73\,k \\
Dielectric tensors & DFPT, PBE & 360\,k \\
Electronic transport & BoltzTraP2 & 1.6\,M \\
Electron--phonon, \Tc & DFPT, PBEsol & 145\,k \\
Phonons; elastic & MLIP & 1.2\,M; 1.7\,M \\
Off-equilibrium & PBE (MatPES) & 429\,k \\
Candidate structures & MLIPs & $>$400\,M \\
\end{tabular}
\end{ruledtabular}
\end{table}

\begin{table}[t]
\caption{Models available to the agent, with mean absolute errors on held-out test sets (median relative error for \epsinf). For
Tuga-SP the two columns give the error from the structure alone (S) and with the orbital-projected
density of states (S + pDOS) of a PBE calculation. Energies in meV/atom, gaps in eV, $S_\textrm{vib}$ in
$k_\textrm{B}$/atom, \Tc\ and $\omega_\textrm{log}$ in K, moduli in GPa.}
\label{tab:models}
\begin{ruledtabular}
\begin{tabular}{@{}llrr@{}}
& & \multicolumn{2}{c}{Error} \\
Model & Property & S & S + pDOS \\
\colrule
Matra-Genoa & crystal structures & -- & -- \\
Tuga-FF & energy, 0\,GPa & 13.6 & -- \\
        & enthalpy, 150\,GPa & 21 & -- \\
Tuga-SP & $E_\textrm{form}$ & 5.8 & -- \\
        & $E_\textrm{hull}$ & 8.6 & -- \\
        & TB-mBJ gap & 0.17 & 0.08 \\
        & \epsinf & -- & 0.8\% \\
        & $S_\textrm{vib}$ (300\,K) & 0.05 & -- \\
        & \Tc & 1.35 & 1.11 \\
        & $\lambda$; $\omega_\textrm{log}$ & 0.084; 25 & 0.067; 22 \\
        & $B$; $G$; $E$ & 7.5; 9.8; 21 & 7.3; 9.0; 20 \\
\end{tabular}
\end{ruledtabular}
\end{table}

\subsection{The Alexandria database}

Table~\ref{tab:alexandria} lists the content of the Alexandria
database~\cite{Schmidt2022,Schmidt2023,Cavignac2025}. For the agent it has three uses. First, it is the reference for stability: convex hulls
with PBE at MatPES and at Materials Project settings, with PBEsol and SCAN, and at pressures up to
150\,GPa, so that the distance of any compound to the hull, and the phases it decomposes into, is
one query away. Second, it is a source of properties computed with DFT, such as TB-mBJ band gaps,
dielectric tensors, transport coefficients and electron--phonon calculations with superconducting
critical temperatures~\cite{Cerqueira2024}. Third, it is a pool of candidates: more than 400
million structures relaxed with machine-learned potentials, which the agent can screen before any
DFT calculation is done. The agent reads a short description of each of these datasets, and of
each of their columns, and writes its own queries.

\subsection{Machine-learning models}

Three families of models trained on these data are available to the agent
(Table~\ref{tab:models}). Matra-Genoa, a transformer generative model, writes new crystals conditioned by the agent
input~\cite{DeBreuck2026}: space group, elements, stoichiometry and Wyckoff positions. Tuga-FF, a
machine-learned interatomic potential trained on Alexandria data computed with MatPES settings at
ambient and high pressure~\cite{DeBreuck2026b}, relaxes a structure in seconds. On the WBM-MatPES
benchmark it predicts energies to 13.6\mevat\ at ambient pressure and enthalpies to 21\mevat\ at
150\,GPa~\cite{DeBreuck2026b}. The graph neural networks of Tuga-SP~\cite{Pires2026} predict
several properties, such as the formation energy and the distance to the hull, the band gaps at the
TB-mBJ level, the electronic dielectric constant, the vibrational entropy, the superconducting
\Tc\ and the elastic moduli. Most of them work from the structure alone. When a PBE calculation of
the structure is available, they also read its orbital-projected density of states~\cite{Pires2026}.
For the TB-mBJ gap this halves the error (Table~\ref{tab:models}). The descriptions of the models give these errors to
the agent, and in the examples below it uses them to decide how far a prediction can be trusted.

\begin{figure*}[t]
    \centering
    \includegraphics[width=\textwidth]{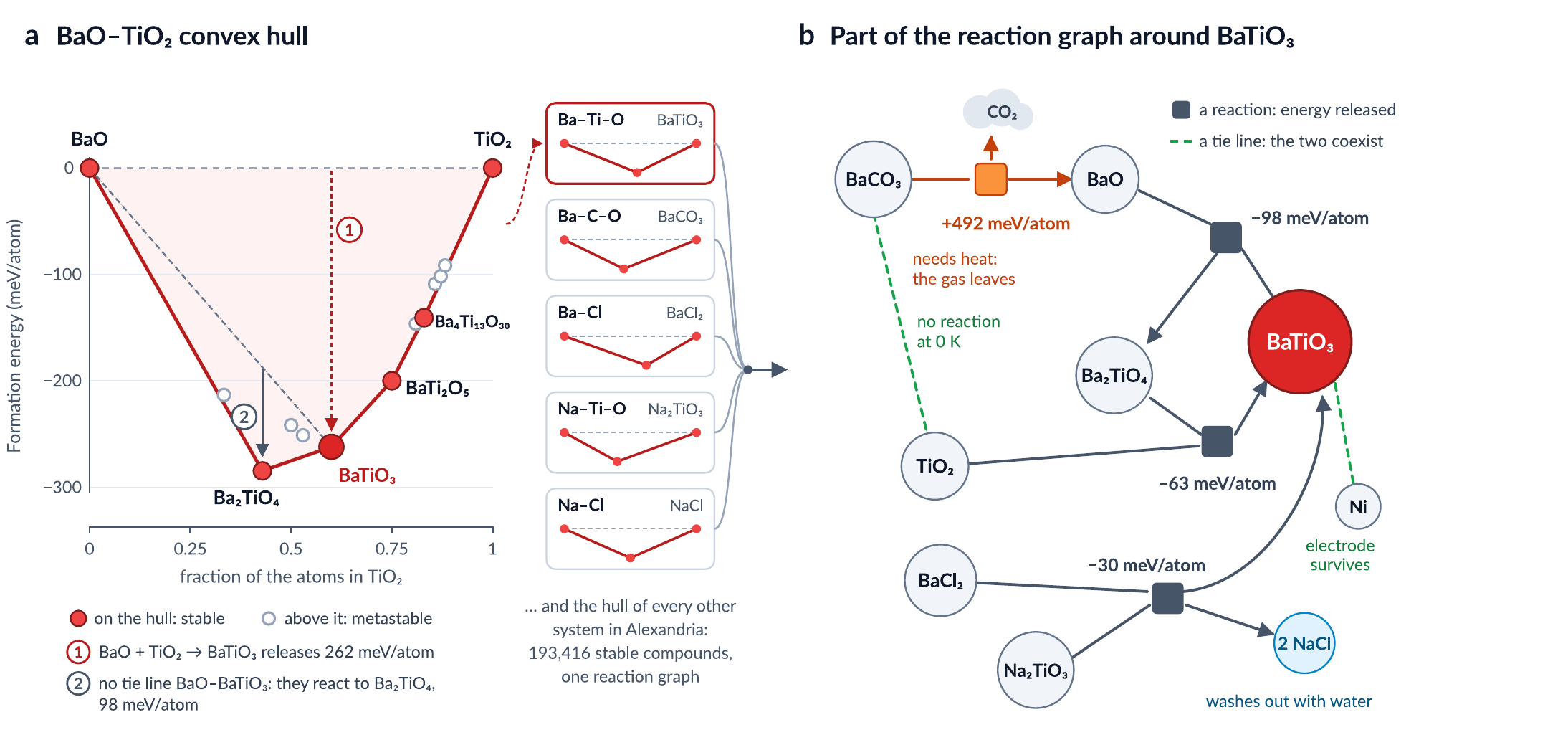}
    \caption{The reaction graph for one compound, \ce{BaTiO3}. (a) The convex hull along the
    \ce{BaO}--\ce{TiO2} line, from the Alexandria energies: filled circles are stable compounds,
    open circles metastable ones. (b) Part of the bipartite graph around \ce{BaTiO3}. Circles are
    compounds, squares reactions, labelled with the energy they release in meV/atom; dashed
    green lines mark pairs of compounds that coexist, as they are not connected through any
    reaction. The carbonate route runs through the
    intermediate \ce{Ba2TiO4}, because \ce{BaO} and \ce{BaTiO3} cannot coexist. Numbers in (b) are
    from the construction of the graph (see \nameref{sec:methods}).}
    \label{fig:graph}
\end{figure*}

\subsection{The reaction graph}
\label{sec:reactionnet}
The convex hull of thermodynamic stability tells us whether a compound is stable and, if not, into
which compounds it decomposes. For a stable compound it also gives the phases with which it can
coexist. To plan a synthesis, however, one needs to know what happens when two given solids are
brought into contact. In principle, the phase diagram contains this information, but only for pairs
of compounds that belong to the same chemical system. In practice, a precursor, a crucible, a flux
and the atmosphere form a system with many elements, for which phase diagrams are rarely computed.

To answer this question we built a reaction graph. This is a bipartite graph with two kinds of
nodes, compounds and reactions, in which each reaction is connected to its reactants and to its
products. Two compounds that are not connected through a reaction can coexist in a stable mix.
Metastable compounds can also be included in this graph,
as nodes that decompose without a second reactant, releasing their distance to the convex hull. In
the same way, a reaction can have three or more reactants. In this work we computed all reactions
between pairs of the 193,416 compounds on the Alexandria hull. For each of the
$1.87\times10^{10}$ pairs we solved a small linear optimisation problem, which finds the set of stable phases
of lowest energy that can be formed from the two compounds while conserving the amount of each
element (Lavoisier's law). Since the products always lie on the hull, a reaction can only run downhill. Of
all pairs, 17.4 billion react and 1.3 billion coexist. The calculation of the complete table took
15,533 core-hours. Reactions with more than two reactants are not stored, but the agent can obtain
them when needed from the same solver. This is the case, for example, of the route with
four reagents in the third example.

Figure~\ref{fig:graph} shows this for barium titanate. Along the
\ce{BaO}--\ce{TiO2} line (Fig.~\ref{fig:graph}a) the hull says that \ce{BaTiO3} is stable and
that \ce{BaO + TiO2 -> BaTiO3} releases 262\mevat.
From the reaction graph (Fig.~\ref{fig:graph}b) we can extract more information. \ce{BaCO3} and \ce{TiO2}, two usual reagents in industrial processes, are joined by a tie line. At
0\,K they do not react at all, and the reaction starts only once the carbonate gives off \ce{CO2},
which needs the entropy of the gas and therefore heat. During heating,
the carbonate decomposes into \ce{BaO} and \ce{CO2}, and the \ce{BaO} formed in this way can be in
contact with \ce{BaTiO3} that has already formed. These two compounds do not coexist: they react to
give \ce{Ba2TiO4}, releasing 98\mevat\ (Fig.~\ref{fig:graph}b). This is the intermediate that
diffraction studies of the carbonate route report, and here it follows from a single reaction of
the graph. \ce{Ba2TiO4} then takes up the remaining \ce{TiO2}
to give \ce{BaTiO3}. The graph also contains the metathesis \ce{BaCl2 + Na2TiO3 -> BaTiO3 + 2 NaCl},
which McDermott \etal\ found to give \ce{BaTiO3} faster and cleaner than the carbonate
route~\cite{McDermott2023}. In the same family it also finds stronger reactions that have not been
tried yet. For instance, \ce{BaI2 + Cs2TiO3 -> BaTiO3 + 2 CsI} releases 203\mevat, almost seven times more, giving \ce{CsI}, which washes out with water, as a side product.

The graph can also be searched exhaustively. Of the 23.4 million pairs
of stable compounds that contain barium, titanium and enough oxygen, 688,006 put all the barium
and titanium into \ce{BaTiO3}. Only 284 of these leave a byproduct that a laboratory can remove, like a
gas, a metal that boils off or a salt that washes out. Six leave nothing at all, and four of the
six are known routes. Of the other two, one rests on an error in the database that the agent was able to identify.

We also need to know how far the predictions of the graph can be trusted. For this
purpose we compared them with the solid-state recipes text-mined from the literature by Kononova
\etal~\cite{Kononova2019} (see \nameref{sec:methods}). Most of the targets and of the precursors lie
on the Alexandria hull or very close to it. However, the comparison also reveals two limitations. The first concerns
reactions that release a gas, such as \ce{BaCO3 + TiO2}. Almost all of these are uphill
at 0\,K, and their energy is meaningless until the entropy of the gas is included. The second is
that ranking the routes by the energy they release does worse than chance in recovering the
precursors chosen by experimentalists. This is not surprising, as the reactions with the largest
driving force involve reactive metals, carbides or peroxides that nobody handles in air. We
therefore see the graph as a generator of candidates, and not as a judge of their quality.
Considerations about handling, availability, hazards and precedent have to come from elsewhere, and
this is what a language model with access to the literature can supply.

The tools built on the graph let the agent ask a few plain questions. Which pairs of compounds
give a target, and what do they leave behind? What does a compound do in contact with typical crucibles, fluxes, electrodes or gases? Over which range of oxygen, sulfur or nitrogen chemical potential
does it survive, and where does air at a given temperature fall in that range? Which reagents are
sold, and how hard are they to handle? How far are computed energies from calorimetry for this
class of compound? In this way, the language model can evaluate the routes proposed by the graph and
give answers that are useful to an experimentalist.

\subsection{Literature and group knowledge}

Finally, we gave the agent access to information that cannot be obtained from a calculation. The agent can search and read the full text
of papers. It can also consult a table of facts on compounds, such as melting and boiling points,
whether a compound is sold, and the price and toxicity of its elements, each with its source, and a
record of the candidates already examined, so that a rejected candidate is not proposed twice. We also wrote short
procedures, for instance how to assess a target compound and how to check the literature, and
numbered rules of thumb that the tools cite in their answers. We are also building a knowledge base for our group,
with its publications, theses and internal documents, which gives an agent the context that a new
member of the group needs months to acquire. The more a research group documents its experience in this form, the better the agent
works for that group.

\begin{figure*}[t]
    \centering
    \includegraphics[width=\textwidth]{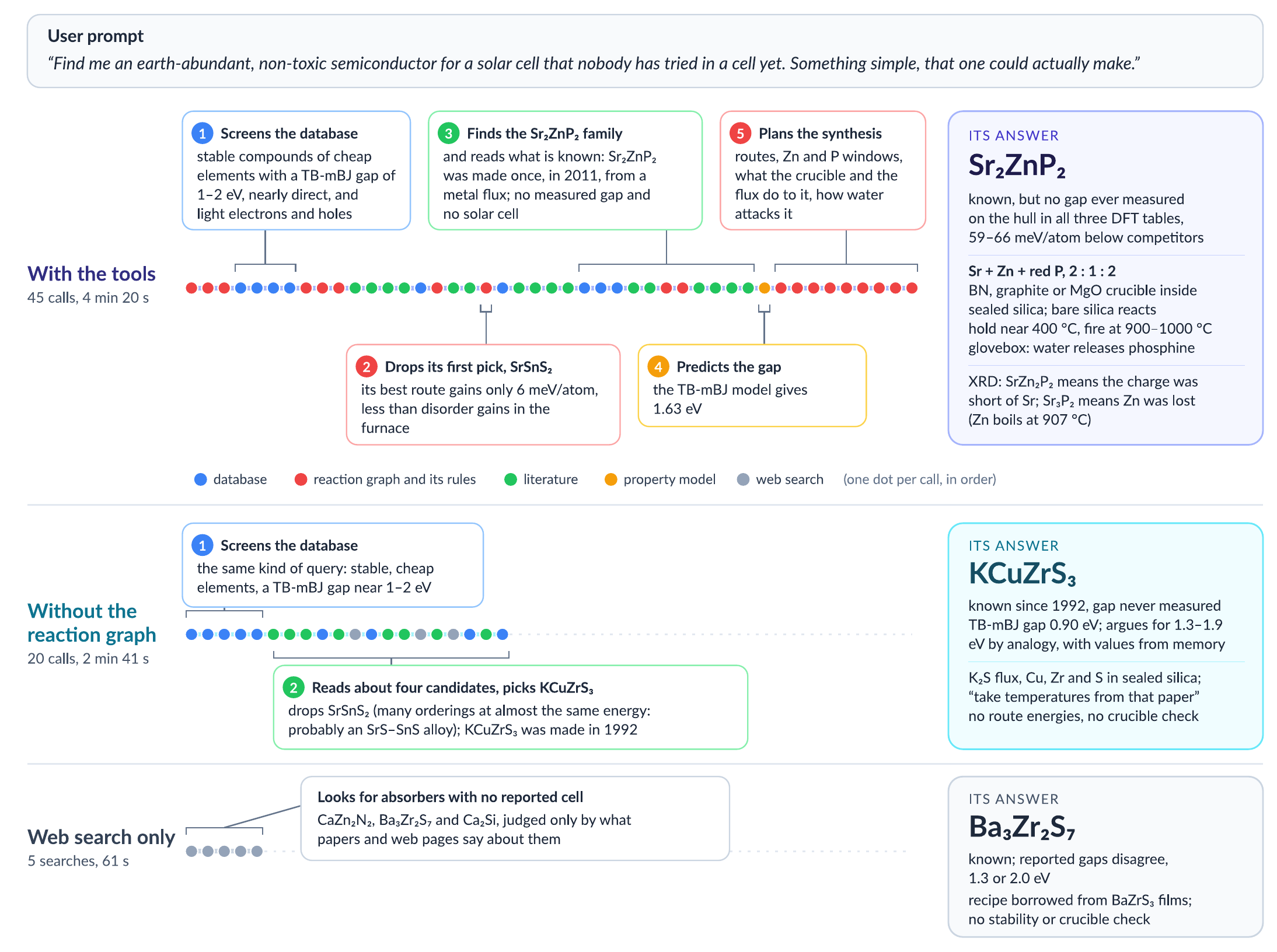}
    \caption{The first example of how our agent works, with three sets of tools. The same model received the same question three times: with all our tools and a web search, with all tools except the reaction graph, and with web search only.
    Each dot is one tool call, in the order the agent made it, coloured by the resource it used; the
    boxes summarise what the agent did at that point, in its own terms. Times are those
    reported by the client. The three transcripts are given in full in the Supplementary
    Information.}
    \label{fig:example}
\end{figure*}

\subsection{Examples}

To show how the agent works with these tools, we used Claude Code with the model Claude Opus 5.5,
unattended in an empty folder, and gave it one plain question each time. For comparison, we asked
the same question to the same agent with other sets of tools: once with all our tools except the
reaction graph, and once with web search only. In the third example we continued the conversation
of the first with two more messages, as a researcher would do. Each answer comes from a single
run. These are therefore examples of what the system does, and not statistics. The full
transcripts are given in the Supplementary Information.

\paragraph*{Example 1: Finding solar absorbers.}
The first question asked for an earth-abundant, non-toxic semiconductor for a solar cell that
nobody has tried in a cell yet, and that one could actually make (Fig.~\ref{fig:example}). These
conditions were chosen on purpose to exclude the established absorbers. Silicon, CdTe,
Cu(In,Ga)Se$_2$ and the lead halide perovskites have all been used in cells, and the last three
also contain toxic or scarce elements (Cd, Te, In, Ga, Pb). The same holds for many of the
alternative absorbers studied over the last two decades. The agent therefore had to look beyond
the materials that dominate the literature.

\begin{figure}[t]
    %\begin{tabular}{c c}
    \centering
    \includegraphics[width=\columnwidth]{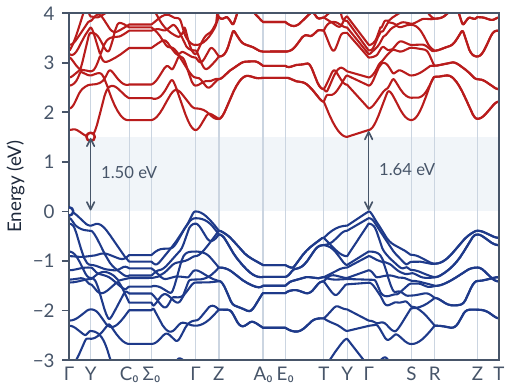} 
    %\includegraphics[width=\columnwidth]{img/fig_SrZrCu2S4_hse_band.pdf} \\
    %(a) & (b) & (c) \\
    %\end{tabular}
\caption{Band structure of \ce{Sr2ZnP2} computed with TB-mBJ, after the agent proposed it. Energies are relative to the top of the valence band (blue); the conduction bands are in red. The gap is indirect, from $\Gamma$ to Y, and the direct gap at $\Gamma$ is 1.64\,eV.}
    \label{fig:sr2znp2}
\end{figure}

When allowed to use all the tools, the agent made 45 tool calls in 4 minutes and 20 seconds. Each
call is a request to one of the tools, such as a database query, a model prediction or a literature
search. It
began by reading the procedure for assessing a target with our tools. It then
wrote its own database query: stable compounds of cheap, non-toxic elements with a band gap
between 1 and 2\,eV, a nearly direct gap and light electrons and holes. For the gap it chose the
TB-mBJ values of the database, a functional known to yield accurate band
gaps~\cite{Tran2009,Borlido2019}, and not the PBE ones. Its first pick,
\ce{SrSnS2}, has a direct gap of 1.6\,eV and is stable in air, but the reaction graph showed that
its best synthesis, \ce{SrS + SnS}, releases only 6\mevat. The agent dropped it, because at furnace
temperatures the entropy of mixing on a shared site is larger than that, so the product would most
likely be a disordered (Sr,Sn)S alloy rather than the ordered compound. It then followed a family
of structures in the database to \ce{Sr2ZnP2}. The literature showed that this compound was made
once, in 2011, from a metal flux, with half of the zinc sites empty~\cite{Wilson2011}; the agent
found no measurement of its band gap and no solar cell. The database does not include a TB-mBJ calculation for \ce{Sr2ZnP2}. To estimate its gap, the agent therefore ran the property model, which predicted a TB-mBJ gap of 1.63\,eV, against 0.80\,eV from PBE.
We later computed the band structure of \ce{Sr2ZnP2} with TB-mBJ (Fig.~\ref{fig:sr2znp2}). The gap
is indirect, 1.50\,eV, and the direct gap at $\Gamma$ is only 0.14\,eV higher. The prediction of the
model is therefore within its expected error (Table~\ref{tab:models}). A hybrid-functional
calculation (HSE06) gives a slightly smaller gap at $\Gamma$, 1.53\,eV against 1.64\,eV. A gap of
1.5--1.6\,eV is somewhat above the optimum for a single-junction cell, about
1.34\,eV~\cite{Shockley1961,Ruhle2016}.

To synthesize \ce{Sr2ZnP2}, the agent used the reaction graph tools and proposed the elements in the ratio
2\,:\,1\,:\,2; a crucible of \ce{BN}, graphite or \ce{MgO} inside a sealed silica tube, because bare silica
and alumina react with the product (by 86 and 32\mevat); and a hold near 400\,\textdegree C to
consume the phosphorus slowly before firing at 900--1000\,\textdegree C. It noted that the compound reacts with water (by 223\mevat) and releases phosphine, so it
should be handled in a glovebox and a solar cell would need encapsulation. It also listed the
secondary phases that can appear in the X-ray pattern. \ce{SrZn2P2} in the product means the charge was missing
strontium, and \ce{Sr3P2} means that zinc was lost, since zinc boils at 907\,\textdegree C.
It also stated clearly that the empty zinc sites of the real compound may change the gap
predicted for the ordered structure. In this
run the agent did not call the generator or the force field since the database already contained the
structures it needed.

In a second run we removed the tools associated with the reaction graph, which also removes the
written procedures, the rules of thumb, the record of examined candidates and the table of compound facts, and
left the database, the models, the literature and the web. The agent made 20 tool calls in 2 minutes
and 41 seconds. It wrote the same kind of database query, examined four candidates,
and dropped \ce{SrSnS2} for a different reason. It observed that several orderings of Sr and Sn lie
at almost the same energy, so the compound is probably a disordered SrS--SnS alloy. It then
proposed \ce{KCuZrS3}, made in 1992~\cite{Mansuetto1992} and on the Alexandria hull, for which it
found no measured band gap. The database gives a direct TB-mBJ gap of only 0.90\,eV; the agent
argued that the real gap is 1.3--1.9\,eV by analogy with copper sulfides whose measured gaps it
quoted from memory, and it said so. The main difference, however, is in the synthesis. Its plan was a potassium
sulfide flux with copper, zirconium and sulfur in sealed silica, with the temperatures to be taken
from the 1992 paper, which it could not open. It gave no energy for any route, no computed check
of the crucible, the flux or the atmosphere (it flagged moisture by analogy with a related
compound), and no guide to the X-ray pattern. The database and the models were
enough to find a good candidate. However, without the reaction graph, and the procedures and facts
that come with it, the agent could not propose a reliable synthesis.

Finally, we gave the agent web search only. It searched five times and after one minute proposed \ce{Ba3Zr2S7}, a
Ruddlesden--Popper sulfide grown as crystals in 2019~\cite{Niu2019}, whose band gap, it reported,
differs between studies (about 1.3 or 2.0\,eV). Its choice and its recipe, adapted from films of \ce{BaZrS3},
rest on what papers say about these compounds. It had no means to check stability, competing
phases or what the film would react with.

\begin{figure*}[t]
    \centering
    \includegraphics[width=\textwidth]{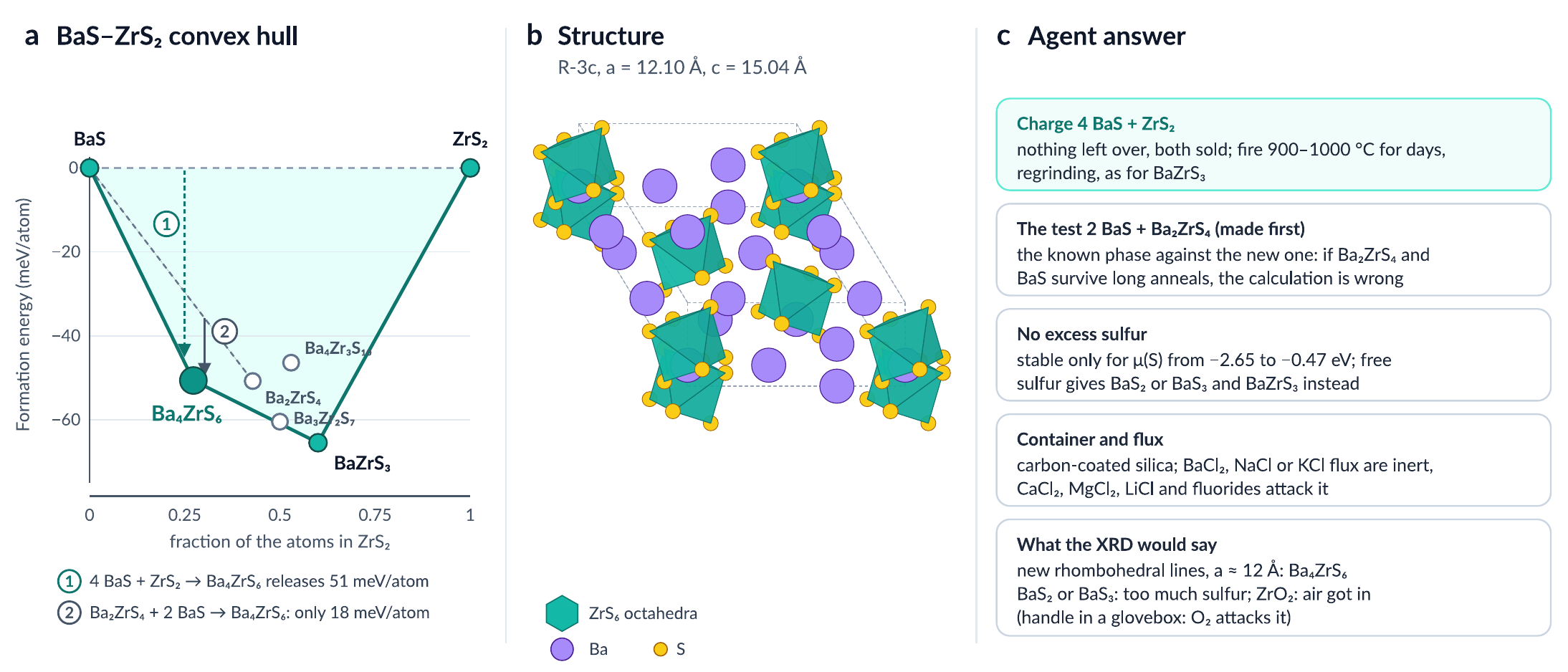}
    \caption{An example of finding a missing phase in a system. (a) The hull along
    the \ce{BaS}--\ce{ZrS2} line, from the same Alexandria energies the reaction graph uses. The
    predicted \ce{Ba4ZrS6} lies on the hull and pushes the known \ce{Ba2ZrS4} 7\mevat\ above it.
    (b) \ce{Ba4ZrS6} structure as relaxed in the database. (c) The synthesis plan of the agent,
    summarised from its answer.}
    \label{fig:bazrs}
\end{figure*}

\paragraph*{Example 2: Expanding known systems.}
The second question was shorter: is there another compound hiding in the Ba--Zr--S system, and if
so, how would one make it? This system is well studied because \ce{BaZrS3} is a candidate solar
absorber, and the known ternary compounds are the perovskite and its Ruddlesden--Popper
relatives~\cite{Niu2019,Kayastha2023}. The agent made 29 tool calls in 2 minutes and 40 seconds. It found in the database \ce{Ba4ZrS6}, a compound of the \ce{K4CdCl6} type with
isolated \ce{ZrS6} octahedra (Fig.~\ref{fig:bazrs}b), which lies 21\mevat\ below its
competitors, and found nothing on it in the literature or on the web. After this analysis it noticed that \ce{Ba4ZrS6} pushes the known compound \ce{Ba2ZrS4}
off the hull by 7\mevat, so that the calculation contradicts an experiment. However, the reaction
deciding between the two, \ce{Ba2ZrS4 + 2 BaS -> Ba4ZrS6}, releases only 18\mevat. That is well
below the typical error of 83\mevat\ that the reaction graph tools report between computed and measured
formation energies of sulfides (Fig.~\ref{fig:bazrs}a).
The agent therefore proposed this reaction as the test that decides between the calculation and
the experiment.
As the direct route, the agent then proposed \ce{4 BaS + ZrS2 -> Ba4ZrS6}, which releases 51\mevat\ and leaves no residual compounds (Fig.~\ref{fig:bazrs}c). The agent also used the reaction graph to choose the conditions. It advised against
excess sulfur, because \ce{Ba4ZrS6} is stable only in a limited range of sulfur chemical potential.
With free sulfur the reaction gives barium polysulfides and \ce{BaZrS3} instead. As container it
proposed a carbon-coated silica ampoule. As flux it proposed \ce{BaCl2}, \ce{NaCl} or \ce{KCl}, but
not \ce{CaCl2}, \ce{MgCl2}, \ce{LiCl} or fluorides, which react with \ce{Ba4ZrS6}. Since the compound
also reacts with oxygen, it should be handled in a glovebox.

\begin{figure*}[t]
    \centering
    \includegraphics[width=\textwidth]{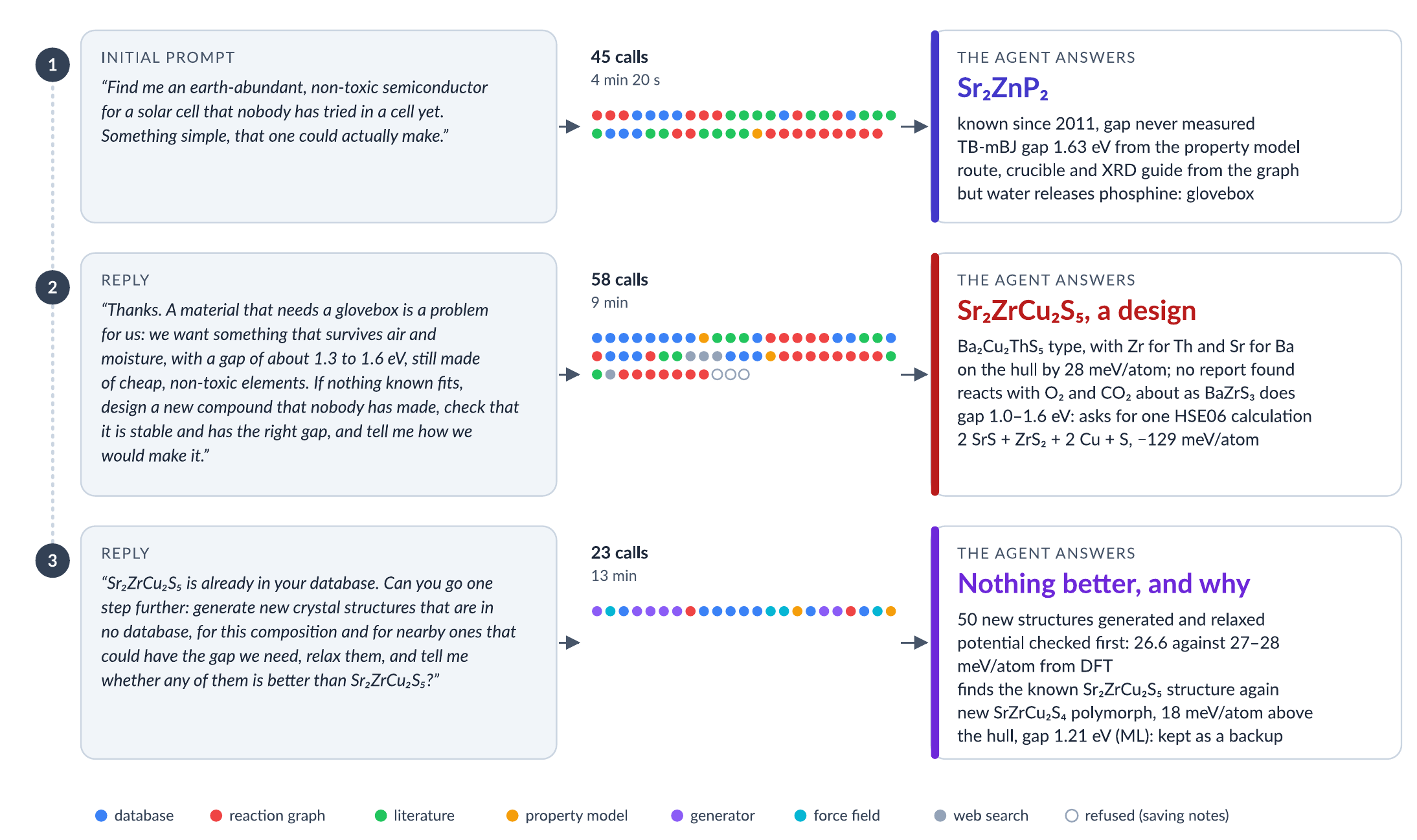}
    \caption{An example of an interactive conversation. The question of the first example
    (Fig.~\ref{fig:example}) is followed by two more messages in the same session, written as a
    researcher would write them. Each dot is one tool call of the agent, in order, coloured by the
    resource it used; open circles are calls that were refused. Times are those reported
    by the client for each message.}
    \label{fig:conversation}
\end{figure*}

\paragraph*{Example 3: Continuing the conversation.}
We then continued the conversation of the first example with two more messages, as a researcher
would (Fig.~\ref{fig:conversation}). The second message noted that a compound that needs a glovebox
is a problem. It asked for one that survives air and moisture, with a gap of about 1.3 to
1.6\,eV and made of cheap, non-toxic elements. If no known compound fitted, the agent should
design one. The agent made 58 tool calls in 9 minutes. It first ruled out the known candidates,
each for a stated reason: already tried in cells, toxic, unstable in air, or failing on its
numbers. It then designed a compound by substitution. \ce{Ba2Cu2ThS5} is a layered sulfide with a
measured gap of 1.86\,eV~\cite{Mesbah2013}. Zirconium in place of thorium should lower the gap,
and strontium in place of barium makes the compound cheaper. The result, \ce{Sr2ZrCu2S5}, is in
the database. It is on the convex hull, 28\mevat\ below its decomposition products, and the agent found no report of it. Its
reactions with oxygen and carbon dioxide release about as much energy as those of \ce{BaZrS3},
which survives air up to 550\,\textdegree C~\cite{Nitta1970}. For water, however, the agent did
not trust the reaction graph. The graph predicts that \ce{SrS} does not react with water, while in
reality it hydrolyses. The agent therefore declared the stability in moisture as untested, and
proposed a simple soak test. It could not settle the band gap either. It placed it between 1.0 and
1.6\,eV, and asked for one hybrid-functional (HSE06) calculation before any experiment. The
synthesis came again from the reaction graph. The charge \ce{2 SrS + ZrS2 + 2 Cu + S} gives only
the target and releases 129\mevat. Excess sulfur gives \ce{CuS} and \ce{ZrS3}, and \ce{NaCl} or
\ce{KI} can serve as a flux.

The third message noted that \ce{Sr2ZrCu2S5} is already in the database, and asked the agent to
look for new structures, for this composition and nearby ones. Here the agent used the generator
and the force field. It generated 50 structures of \ce{Sr2ZrCu2S5}, \ce{SrZrCu2S4} and
\ce{Ca2ZrCu2S5}, and relaxed them. Before trusting the force field, it checked that it reproduces
the DFT stability of the known compounds, to within about 10\mevat. For \ce{Sr2ZrCu2S5} the
generator found again the structure of the database, and nothing better. However, it found a new
polymorph of \ce{SrZrCu2S4}, close to the hull and much more stable than the only structure of
this composition in Alexandria. The property model predicted a TB-mBJ gap of 1.21\,eV for this
polymorph, against 0.68\,eV for \ce{Sr2ZrCu2S5}. The agent kept \ce{Sr2ZrCu2S5} as the target. It
offered the polymorph as a backup, in case the hybrid calculation gives too small a gap, and
stated that nothing in this round had been checked with DFT.

%The caution of the agent was justified. We later computed the band structure of
%\ce{Sr2ZrCu2S5} with TB-mBJ and found an indirect gap of only 0.40\,eV, too small for a solar
%absorber. The polymorph of \ce{SrZrCu2S4}, with a calculate TB-mBJ gap of 0.94~eV, is therefore the better candidate.

The caution of the agent was justified. We later computed the band structures of both compounds with HSE06. \ce{Sr2ZrCu2S5} has an indirect gap of only 0.98\,eV, from $\Gamma$ to Y$_2$, which is too small for a solar absorber. The new polymorph of \ce{SrZrCu2S4} has a direct gap of 1.63\,eV at $\Gamma$ and is therefore the better candidate.

This is only a small example. The three messages took the agent 26 minutes in total. A
real conversation would continue until the researcher is satisfied with the proposal and it has been validated. It could then turn to doping and defects, to the band alignment with contact
layers, using the branch-point energies already in the database, or to the lattice matching with
a substrate. The example shows only the first steps of such a conversation.

%=======================================================================
\section{Discussion}
\label{sec:discussion}
%=======================================================================

We connected a general-purpose agent to the data and models of our group and to a reaction graph
that covers the whole Alexandria hull, and let it work unattended. We offer these results as a
perspective on what awaits materials science in the coming years.

Most agents proposed so far for materials science give the model the means to run calculations:
DFT codes, interatomic potentials and generative
models~\cite{Kumar2026,Wang2025,YangP2026,Ghafarollahi2026,Deng2026}. However, a frontier model can
already run such codes by itself. A new calculation also answers only a few of the questions of
an experimentalist. What the model lacks are the data that already exist, and a way to reason
about synthesis. In our runs the same model behaved very differently depending on what
it could access. With web search alone it proposed what the literature already discusses and
borrowed recipes from related compounds. With the database and the models but no reaction graph it
found and checked candidates, but its synthesis advice still came from the literature. Staley
\emph{et al.}~reached a similar conclusion for the Nb--O system~\cite{Staley2026}. In their work,
the prior knowledge of a language model, combined with a thermodynamic simulation, gave more viable
routes than a classical path search alone. Machine-learned synthesizability scores, in contrast, often
disagree with thermodynamics~\cite{Schlesinger2026}, which is one more reason to give the agent the
energies themselves.

With all the tools, the agent took its numbers from the tools and not from its memory. It gave a
reason for every candidate it dropped. Its instructions can be followed in the laboratory, and if
they are wrong, an experiment will show it. Most of what made the answers useful came from
combining tools: a database query to find candidates, a reaction to estimate a margin, a paper to
see what is known, and a contact table to choose the crucible. For a researcher such work can take
days, while the agent needed only a few minutes. The more data and experience of a group
the agent can access, the better its decisions. Documenting past leads and writing down the
procedures of a research group can make the agent much better.

We think this way of working will become a major approach in materials science in the coming years, and that it can bring
computation and experiment closer.
An experimental group can ask a question in plain words, and the
answer is based on millions of calculations. The group does not need to learn a database schema or
a simulation code. Computational groups can work behind the agents, providing the data and models
that make every answer better. For this the community will have to change how it shares results.
Databases with a web page and an interface for programs are a start. Common standards such as
OPTIMADE~\cite{Andersen2021} already allow different databases to be queried in the same way. Agents need more: tools that
answer the questions experimentalists ask, and descriptions that say what each number means and
when it is wrong. Connected to autonomous laboratories~\cite{Szymanski2023,Tom2024,Chen2024}, such
agents could close the loop from a question to a sample. We see them as a path for future research, linking those who compute materials with those who make them.

%=======================================================================
\section{Methods}
\namedlabel{sec:methods}{Methods}
%=======================================================================

\subsection{Agent runs}

All runs used Claude Code 2.1.308 in non-interactive mode, with the model Claude Opus 5.5 at its
default settings. Each run received one question. The only exception is the third example, in
which two more messages were sent in the same session as the first. Each run started in a new,
empty folder, without project files, memory or settings of ours. Besides the question, the only
instructions the agent received were therefore the descriptions of the tools. In the
runs with tools, the agent could call our server and web search. Only the tool that writes to the
table of compound facts was disabled. In the run without the reaction graph, we disabled all 26
tools of the reaction-graph server. This left 16 tools: the database, the three models and the
literature. The procedures, the rules of thumb, the record of examined candidates and the table of
compound facts are on that server, so they were removed as well. In the run with web search only,
the agent could only search the web and fetch pages.

\subsection{Construction of the reaction graph}

The hull is the PBE hull of Alexandria with Materials Project settings, with 193,416 compounds.
Two compounds $A$ and $B$ coexist if no mixture of them can lower its energy by rearranging into
other hull phases. For each pair we solve
\begin{align}
&\min_{y,\,n\,\ge\,0}\; y_A E_A + y_B E_B + \textstyle\sum_i n_i E_i \nonumber\\
&\text{subject to}\quad y_A c_A + y_B c_B + \textstyle\sum_i n_i c_i = c_A + c_B ,
\end{align}
where $E$ and $c$ are energies and compositions and the sum runs over hull compounds made only of
the elements of $A$ and $B$ (a median of 129 candidates). If both $y_A$ and $y_B$ end below one,
the pair reacts. Reactions with more than two reactants are solved in the same way, with one
term $y_k E_k$ for each reactant. Mass balance was checked on 19,997 random reactions and
the stored product sets were recomputed for a sample. Routes to a target fix the ratio of two
precursors so that the cations come in the ratio of the target, and decompose the mixture on the
hull. Gases enter as reservoirs: \ce{H2O} and \ce{CO2} with experimental formation free energies,
\ce{O2} and \ce{N2} with the corrected references of the database; stability windows come from
sweeping one chemical potential at a time. The
comparison with experiment uses the 3,951 usable recipes among the 19,488 text-mined from the
literature~\cite{Kononova2019}. Of the compounds made by solid-state reaction, 59\% sit on the
Alexandria hull and 92\% lie within 50\mevat\ of it, and 98\% of the precursors are on the hull.
Reactions that release a gas are uphill at 0\,K in 95\% of these recipes. Ranking the routes first
by the number of reagents and then by the smallest driving force puts the experimental choice
first in 44\% of 1,510 recipes, and in the top ten in 92.5\%, out of a median of 42 candidates.

\section*{Data availability}

The Alexandria database is available at \url{https://alexandria.icams.rub.de}. The transcripts of
the four runs are given in the Supplementary Information.

\section*{Acknowledgements}

The authors gratefully acknowledge the computing time made available to them on the high-performance computers Noctua and Otus at the NHR Center Paderborn Center for Parallel Computing (PC2).
The authors gratefully acknowledge the scientific support and HPC resources provided by the Erlangen National High Performance Computing Center (NHR@FAU) of the Friedrich-Alexander-Universität Erlangen-Nürnberg (FAU) under the NHR project k114eb. Part of the calculations were performed on the HPC cluster Elysium of the Ruhr University Bochum, subsidised by the DFG (INST 213/1055-1).
The authors also acknowledge the Claude credits and subscriptions given through the Claude Science cohort of Anthropic.

\section*{Funding}

P.R.P., T.C., and M.A.L.M. were supported by a collaboration between the Kavli Foundation, Klaus Tschira Stiftung, and Kevin Wells, as part of the SuperC collaboration, and by the Simons Foundation through the Collaboration on New Frontiers in Superconductivity (Grant No. SFI-MPS-NFS-00006741-10).

\renewcommand{\bibsection}{\section*{References}}
\bibliography{bib.bib}

\end{document}

\begin{center}
{\Large Supplementary Information}\\[6pt]
{\large Materials research with a general-purpose agent}\\[4pt]
Paulo R. Pires, Th\'{e}o Cavignac, Tiago F. T. Cerqueira, Hai-Chen Wang, Silvana Botti and Miguel A. L. Marques
\end{center}

\section*{S1. The tools}

The tools of our server, with the first sentence of the description the agent reads. The full
descriptions are part of the code of the servers.

\begin{longtable}{lp{12.2cm}}
\toprule
tool & first sentence of its description \\
\midrule
\endhead
\texttt{alexandria\_materials\_query} & Run one read-only SQL statement (SELECT or WITH) on the Alexandria database. \\
\texttt{alexandria\_tables} & The  tables of the database, each with the database's own description of it and its number of rows (`rows` counted, `approx\_rows` estimated above 100,000). \\
\texttt{alexandria\_describe\_table} & One table's shape: its number of rows (`rows` counted, or `approx\_rows` from the statistics above 100,000), every column with its type, in order, and the database's comme ... \\
\texttt{alexandria\_find\_structure} & Entries of a compound in Alexandria, most stable first. \\
\texttt{alexandria\_structure\_cif} & One structure as a CIF: the symmetrized conventional cell, with its space group, cell parameters and every inequivalent site (label, Wyckoff position, multiplicity, a fra ... \\
\texttt{alexandria\_view\_structure} & LOOK at a structure: one image with the conventional cell seen down a, b and c, ball-and-stick, partial occupancy as pie slices. \\
\texttt{matra\_generate} & Generate new crystal structures, optionally with a given space group, elements and stoichiometry. \\
\texttt{matra\_info} & The model this server runs, the most structures one call returns, and the GPU memory in use. \\
\texttt{tugaff\_relax} & Relax crystal structures: the atoms, and unless relax\_cell is false the cell too, move until every force is below  eV/A, for at most  steps (BFGS). \\
\texttt{tugaff\_singlepoint} & Energy and forces of crystal structures as they are, without moving anything. \\
\texttt{tugaff\_info} & The model this server runs, the relaxation settings (force threshold, step limit), the structure formats it reads, and the GPU memory in use. \\
\texttt{tugasp\_predict} & Predict properties of crystal structures. \\
\texttt{tugasp\_models} & What every property is (definition, units), which inputs its model reads, how accurate it is on held-out test materials and where it is weak, and how to build the two DOS ... \\
\texttt{tugasp\_health} & Which property models are on the GPU now and how long they have been unused, and the GPU memory in use. \\
\texttt{literature\_search\_papers} & Search the scholarly literature. \\
\texttt{literature\_read\_paper} & Read a paper's full text: a DOI, an arXiv id, a title, or a URL (a PDF, or any web page, which comes back as text). \\
\texttt{reactions\_conventions} & The conventions of this project that silently produce a WRONG ANSWER rather than an error. \\
\texttt{reactions\_design\_rules} & The project's design rules, R1-R28, which the other tools cite by number: R6 ranks routes, R9 is the 25 meV/atom tolerance, R22 the f-electron cross-check, R26 keeps avai ... \\
\texttt{reactions\_leads} & The leads ledger: every candidate that reached a literature check, with its verdict, the reason, the archetype (A1-A12, see guide('campaign')) and the rules it bears on. \\
\texttt{reactions\_guide} & How to use these tools: the procedures, in order, and how to read the answers. \\
\texttt{reactions\_kinetic\_floor} & The temperature bracket a charge can react in, from Tammann's rule: a solid's own lattice barely diffuses below about T\_m/2. \\
\texttt{reactions\_reagent\_profile} & Everything known about one reagent, in one flat answer that starts with a `summary` in words: whether it is a solid a laboratory could weigh out (usable\_reagent, and why\_ ... \\
\texttt{reactions\_compare\_reagents} & reagent\_profile for several formulas -- a shortlist, or the reagents of one route -- plus the weakest link: a route is only as manufactured as its least manufactured reag ... \\
\texttt{reactions\_resolve\_target} & Canonicalise a compound and say where it sits. \\
\texttt{reactions\_decompose} & The equilibrium assemblage at a composition: what this mixture of elements actually becomes at 0 K. \\
\texttt{reactions\_reaction} & What a weighed-out charge goes to, and how far downhill. \\
\texttt{reactions\_balance} & Solve the stoichiometry of reagents -> target (+ n byproduct) exactly, or report that it cannot balance. \\
\texttt{reactions\_margin\_at\_mu} & How far a target sits above the hull when some elements are OPEN at a fixed chemical potential -- the quantity a furnace actually sets, as opposed to the closed-system nu ... \\
\texttt{reactions\_byproduct\_free\_routes} & Routes to the target that leave NOTHING to separate: every combination of up to maxk phases from the target's own chemical system whose composition is exactly the target' ... \\
\texttt{reactions\_reservoir\_routes} & Routes that need a gas reservoir, for a target a closed crucible cannot reach. \\
\texttt{reactions\_atmosphere\_potentials} & What a furnace atmosphere delivers: the chemical potential per atom of each element it sets, in eV, displaced from that element's own 0 K reference -- the axis decompose  ... \\
\texttt{reactions\_firing\_window} & The firing window: over what range of an open element is the target the phase you actually get. \\
\texttt{reactions\_calorimetry} & What a measured formation enthalpy says about a computed one -- the error bar, never a correction. \\
\texttt{reactions\_vehicle\_search} & Which solid DELIVERS a mobile element at the activity the target needs. \\
\texttt{reactions\_contact} & What the target survives contact with: crucibles, atmosphere and molten-salt fluxes. \\
\texttt{reactions\_offstoichiometry} & For a target ABOVE the hull: what an off-stoichiometric stabilisation would have to supply. \\
\texttt{reactions\_metastable\_routes} & For a target above the hull: exact-stoichiometry routes from every computed phase within 'cut' of the hull, plus the invariant margin every one of them must overcome and  ... \\
\texttt{reactions\_screen\_find} & Every row in every screen that mentions a compound -- high pressure, sibling gaps, ion exchange, interlayers, atmosphere, thermites. \\
\texttt{reactions\_screen\_query} & Read the whole-hull screens. \\
\texttt{reactions\_reactions\_with} & One compound against a list of named partners: which it coexists with and which attack it, sorted by how hard. \\
\texttt{reactions\_compound\_data} & Everything on file about one compound, each value with where it came from: the melting, boiling and decomposition points and their source (the same values reagent\_profile ... \\
\texttt{reactions\_add\_compound\_data} & Put one fact about a compound on file, with the DOI or web page it comes from: never from memory, fetch the source first (literature\_search\_papers, literature\_read\_paper) ... \\
\bottomrule
\end{longtable}

\section*{S2. How the agents were run}

Every run used Claude Code 2.1.308 in non-interactive mode, with the model Claude Opus 5.5
(\texttt{claude-opus-5-5}) at its default settings. Each run
started in a new, empty directory, with no project files, memory or settings of ours, so that the
only instructions the agent received beyond the prompt were the descriptions of the tools.

Follows the agent prompt, the setup, the tool calls in the order the agent made them (arguments
shortened to 400 characters, crystal structures replaced by \texttt{[...]}), and the final
answer as the agent wrote it. Characters outside ASCII are spelled out (subscripts as digits,
\texttt{->} for arrows, \texttt{deg} for the degree sign).

\subsection*{first example, solar absorber: web search only}
2 tools available; 5 tool calls; 6 turns; 61\,s.

Prompt:
\begin{Verbatim}[fontsize=\footnotesize]
Find me an earth-abundant, non-toxic semiconductor for a solar cell that nobody has tried in
  a cell yet. Something simple, that one could actually make.
\end{Verbatim}
Tool calls:
\begin{Verbatim}[fontsize=\footnotesize]
 1. WebSearch({"query": "CaZn2N2 thin film solar cell device", "mode": "extended"})
 2. WebSearch({"query": "Ba3Zr2S7 thin film solar cell device fabricated", "mode":
  "extended"})
 3. WebSearch({"query": "Ca2Si thin film photovoltaic device direct band gap", "mode":
  "extended"})
 4. WebSearch({"query": "Ba3Zr2S7 thin films growth photoluminescence Ruddlesden-Popper
  chalcogenide", "mode": "extended"})
 5. WebSearch({"query": "CaZn2N2 synthesis thin film sputtering band gap experimental",
  "mode": "extended"})
\end{Verbatim}
Answer:
\begin{Verbatim}[fontsize=\footnotesize]
I'd go with **Ba3Zr2S7**, a layered sulfide made from barium, zirconium and sulfur. I can't
  prove nobody has built a cell from it, but my searches found no device reports. The
  closest work is on BaZrS3, its better-known perovskite relative, which still has only a
  few crude attempts at cells.

**Why it fits what you asked for**
- **Earth-abundant and non-toxic:** Ba, Zr and S are all cheap and plentiful, with no Cd,
  Pb, Se, Te or In. The one handling concern is that soluble barium salts are toxic, but the
  sulfide itself is a stable ceramic.
- **Simple:** it's a single known compound. Bulk crystals have already been grown by a flux
  method and characterised by photoluminescence and Raman ([J. Mater. Res.](https://www.camb
  ridge.org/core/journals/journal-of-materials-research/article/abs/crystal-growth-and-struc
  tural-analysis-of-perovskite-chalcogenide-bazrs3-and-ruddlesdenpopper-phase-ba3zr2s7/D61F0
  3D50FC1510FF32E9D7E7FA0B93F)).
- **You could actually make it:** the BaZrS3 thin-film methods should carry over. One route
  deposits BaS and Zr in layers, then anneals them in sulfur ([Adv. Eng. Mater.
  2024](https://advanced.onlinelibrary.wiley.com/doi/full/10.1002/adem.202302161)).
  Sputtering followed by a high-temperature anneal also works ([ACS AEM](https://pubs.acs.or
  g/aaemcq/article/6/22/11642/332536/Interplay-between-Growth-Mechanism-Materials)). To
  target Ba3Zr2S7, you mainly shift the Ba:Zr ratio to 3:2. Ba3Zr2S7 already shows up as an
  unwanted extra phase in Ba-rich BaZrS3 films, which suggests it forms readily.
- **Open question worth answering:** the measured band gap disagrees between studies, about
  2.0 eV from absorption versus 1.28 eV from photoluminescence. At 1.3 eV it would work as a
  single-junction absorber; at about 2 eV it would suit the top cell of a tandem stack. A
  first device would help settle which it is.

**Risks you should know about**
- Sulfurization usually needs about 900 degC, so you'd need quartz or another refractory
  substrate, and Mo back contacts may degrade at that temperature.
- Defects that waste charge carriers are a known problem in BaZrS3: some films show no
  room-temperature photoluminescence at all ([J. Mater. Chem. A](https://pubs.rsc.org/ta/art
  icle/doi/10.1039/d5ta05550d/1280108/Effect-of-composition-on-structural-and)). Ba3Zr2S7
  may have the same issue.
- Raman alone can't tell Ba3Zr2S7 apart from Ba4Zr3S10, so you'll need X-ray diffraction to
  confirm which phase you made.
- A sensible first cell would be glass/FTO, then a TiO2 or CdS-free ZnO electron layer, then
  Ba3Zr2S7, then a hole contact such as Spiro or NiO, then Au. Expect a very low efficiency
  at first; just getting a working device would be the result worth publishing.

**Candidates I looked at and rejected**
- **CaZn2N2** has a gap of about 1.9 eV and also appears to have no cells. But only MBE
  (molecular beam epitaxy) has produced films; other physical vapour deposition and
  annealing routes failed ([ACS AEM](https://pubs.acs.org/aaembp/article-abstract/1/8/1433/6
  18701/Heteroepitaxial-Thin-Film-Growth-of-a-Ternary)). That's too hard for "simple".
- **Ca2Si** has a gap of about 1.0-1.2 eV, epitaxial films exist on Si, and I found no cells
  ([Vacuum](https://www.sciencedirect.com/science/article/abs/pii/S0042207X22004262), [Resea
  rchGate](https://www.researchgate.net/publication/346823742_Transport_Properties_of_CaSi2_
  and_Ca2Si_Thin_Films)). However, it reacts with air and moisture, which makes making and
  keeping a device much harder. It's the backup if you have a glovebox and an evaporator.

Before you commit, do a proper search of Web of Science or Scopus and arXiv for "Ba3Zr2S7"
  together with "solar cell" or "photovoltaic". This field moves quickly, and web search can
  miss conference papers and recent preprints.

Sources:
- [Crystal growth of BaZrS3 and Ba3Zr2S7 (J. Mater. Res.)](https://www.cambridge.org/core/jo
  urnals/journal-of-materials-research/article/abs/crystal-growth-and-structural-analysis-of
  -perovskite-chalcogenide-bazrs3-and-ruddlesdenpopper-phase-ba3zr2s7/D61F03D50FC1510FF32E9D
  7E7FA0B93F)
- [Same, ResearchGate](https://www.researchgate.net/publication/337603168_Crystal_growth_and
  _structural_analysis_of_perovskite_chalcogenide_BaZrS_3_and_Ruddlesden-Popper_phase_Ba_3_Z
  r_2_S_7)
- [BaZrS3 stacked-layer films (Adv. Eng. Mater.
  2024)](https://advanced.onlinelibrary.wiley.com/doi/full/10.1002/adem.202302161)
- [Sputtered BaZrS3 films (ACS AEM)](https://pubs.acs.org/aaemcq/article/6/22/11642/332536/I
  nterplay-between-Growth-Mechanism-Materials)
- [Combinatorial BaZrS3 films (JMCA)](https://pubs.rsc.org/ta/article/doi/10.1039/d5ta05550d
  /1280108/Effect-of-composition-on-structural-and)
- [Epitaxial chalcogenide perovskite films (arXiv)](https://arxiv.org/pdf/2105.11090)
- [CaZn2N2 MBE films](https://pubs.acs.org/aaembp/article-abstract/1/8/1433/618701/Heteroepi
  taxial-Thin-Film-Growth-of-a-Ternary)
- [Zintl AM2Pn2 map](https://www.researchgate.net/publication/392976954_Map_of_the_Zintl_AM_
  2_Pn_2_Compounds_Influence_of_Chemistry_on_Stability_and_Electronic_Structure)
- [Ca2Si
  photoconductivity](https://www.sciencedirect.com/science/article/abs/pii/S0042207X22004262
  )
- [Ca2Si/CaSi2 transport](https://www.researchgate.net/publication/346823742_Transport_Prope
  rties_of_CaSi2_and_Ca2Si_Thin_Films)
\end{Verbatim}
\subsection*{first example, solar absorber: our tools and web search}
43 tools available; 45 tool calls; 46 turns; 260\,s.

Prompt:
\begin{Verbatim}[fontsize=\footnotesize]
Find me an earth-abundant, non-toxic semiconductor for a solar cell that nobody has tried in
  a cell yet. Something simple, that one could actually make.
\end{Verbatim}
Tool calls:
\begin{Verbatim}[fontsize=\footnotesize]
 1. reactions_guide({"name": "assess-target"})
 2. reactions_guide({"name": "check-literature"})
 3. reactions_leads({})
 4. alexandria_describe_table({"table": "energy_runs_mbj"})
 5. alexandria_describe_table({"table": "el_transport_pbe"})
 6. alexandria_materials_query({"sql": "SELECT p.mat_id, p.formula, p.spg, p.nsites,
  round(p.band_gap_ind::numeric,2) pbe_gi, round(p.band_gap_dir::numeric,2) pbe_gd,
  round(m.band_gap_ind::numeric,2) mbj_gi, round(m.band_gap_dir::numeric,2) mbj_gd,
  round(t.m_dos_e::numeric,2) me, round(t.m_dos_h::numeric,2) mh\nFROM energy_runs_pbe p
  JOIN energy_runs_mbj m ON m.mat_id=p.mat_id JOIN el_transport_pbe t ON
  t.mat_id=p.mat_id\nWHERE ...)
 7. alexandria_materials_query({"sql": "SELECT p.mat_id, p.formula, p.spg, p.nsites,
  round(p.band_gap_ind::numeric,2) pbe_gi, round(m.band_gap_ind::numeric,2) mbj_gi,
  round(m.band_gap_dir::numeric,2) mbj_gd, round(t.m_dos_e::numeric,2) me,
  round(t.m_dos_h::numeric,2) mh\nFROM energy_runs_pbe p JOIN energy_runs_mbj m ON
  m.mat_id=p.mat_id LEFT JOIN el_transport_pbe t ON t.mat_id=p.mat_id\nWHERE p.e_above_hull
  <= 0.005 AND p.nsite ...)
 8. reactions_resolve_target({"query": "SrSnS2", "all_tables": true})
 9. reactions_leads({"compound": "SrSnS2"})
10. reactions_screen_find({"compound": "SrSnS2"})
11. literature_search_papers({"query": "SrSnS2 strontium tin sulfide", "sources":
  ["openalex", "arxiv", "crossref"]})
12. literature_search_papers({"query": "SrSnS2 crystal structure tin(II)", "sources":
  ["openalex", "crossref"]})
13. literature_search_papers({"query": "SrSnS2 photovoltaic absorber", "sources":
  ["openalex", "arxiv"]})
14. literature_search_papers({"query": "BaSnS2 tin(II) thiostannate solar absorber band
  gap", "sources": ["openalex", "arxiv"]})
15. alexandria_structure_cif({"structure": "agm096925532"})
16. reactions_reagent_profile({"formula": "SrSnS2"})
17. literature_search_papers({"query": "SnS SrS solid solution rocksalt alloy", "sources":
  ["openalex", "arxiv"]})
18. literature_search_papers({"query": "alkaline earth tin(II) sulfide CaSnS2 BaSnS2
  synthesis", "sources": ["openalex", "crossref"]})
19. reactions_byproduct_free_routes({"target": "SrSnS2", "top": 10})
20. alexandria_materials_query({"sql": "SELECT p.mat_id, p.formula, p.spg, p.nsites,
  round((p.e_phase_separation*1000)::numeric,0) depth_meV, round(p.band_gap_ind::numeric,2)
  pbe_gi, round(m.band_gap_ind::numeric,2) mbj_gi, round(m.band_gap_dir::numeric,2) mbj_gd,
  round(t.m_dos_e::numeric,2) me, round(t.m_dos_h::numeric,2) mh\nFROM energy_runs_pbe p
  JOIN energy_runs_mbj m ON m.mat_id=p.mat_id LEFT JOIN el_transport_pbe t ON t.m ...)
21. literature_search_papers({"query": "BaZrN2 nitride semiconductor thin film
  photovoltaic", "sources": ["openalex", "arxiv"]})
22. literature_search_papers({"query": "Ba2MgP2 phosphide Zintl", "sources": ["openalex",
  "arxiv", "crossref"]})
23. literature_search_papers({"query": "Ca2ZnP2 phosphide", "sources": ["openalex", "arxiv",
  "crossref"]})
24. literature_search_papers({"query": "Na2CuP sodium copper phosphide", "sources":
  ["openalex", "arxiv", "crossref"]})
25. alexandria_materials_query({"sql": "SELECT p.mat_id, p.formula, p.prototype_id,
  round((p.e_above_hull*1000)::numeric,0) eah, round((p.e_phase_separation*1000)::numeric,0)
  depth, round(m.band_gap_ind::numeric,2) mbj_gi, round(m.band_gap_dir::numeric,2)
  mbj_gd\nFROM energy_runs_pbe p LEFT JOIN energy_runs_mbj m ON m.mat_id=p.mat_id\nWHERE
  p.prototype_id IN (SELECT prototype_id FROM energy_runs_pbe WHERE mat_id IN ('agm0022858
  ...)
26. alexandria_materials_query({"sql": "SELECT p.mat_id, p.formula,
  round((p.e_phase_separation*1000)::numeric,0) depth, round(p.band_gap_ind::numeric,2)
  pbe_gi, round(p.band_gap_dir::numeric,2) pbe_gd, round(m.band_gap_ind::numeric,2) mbj_gi,
  round(m.band_gap_dir::numeric,2) mbj_gd, round(t.m_dos_e::numeric,2) me,
  round(t.m_dos_h::numeric,2) mh\nFROM energy_runs_pbe p LEFT JOIN energy_runs_mbj m ON
  m.mat_id=p.mat_id LEFT JOIN  ...)
27. alexandria_structure_cif({"structure": "agm002232584"})
28. literature_search_papers({"query": "Sr2ZnP2", "sources": ["openalex", "arxiv",
  "crossref"]})
29. literature_search_papers({"query": "strontium zinc phosphide Sr2ZnP2 crystal structure",
  "sources": ["openalex", "crossref"]})
30. reactions_leads({"compound": "Sr2ZnP2"})
31. reactions_screen_find({"compound": "Sr2ZnP2"})
32. literature_read_paper({"identifier": "10.1002/zaac.201100177", "find": "band
  gap,resistivity,semiconduct,air,moisture,flux,optical,color,colour,Sr2ZnP2", "max_chars":
  20000})
33. literature_search_papers({"query": "Sr2ZnP2 solar cell absorber optical band gap",
  "sources": ["openalex", "arxiv"], "year_from": 2012})
34. literature_search_papers({"query": "A2ZnP2 Zintl phosphide photovoltaic Sr2ZnP2 Eu2ZnP2
  Ca2ZnP2", "sources": ["openalex", "arxiv"]})
35. literature_read_paper({"identifier": "arXiv:2310.18188", "find":
  "Sr2ZnP2,SrZn,Zn0.5,partial occupan,disorder,candidates,Table", "max_chars": 15000})
36. tugasp_predict({"structures": "[...]", "properties": ["mbj_gap", "band_gap_ind"]})
37. reactions_design_rules({"ids": ["R6", "R7", "R9", "R26"]})
38. reactions_resolve_target({"query": "Sr2ZnP2", "all_tables": true})
39. reactions_byproduct_free_routes({"target": "Sr2ZnP2", "top": 10})
40. reactions_firing_window({"target": "Sr2ZnP2", "element": "Zn"})
41. reactions_firing_window({"target": "Sr2ZnP2", "element": "P"})
42. reactions_contact({"target": "Sr2ZnP2", "groups": ["metals", "ceramics", "atmosphere",
  "fluxes"]})
43. reactions_compare_reagents({"formulas": ["Sr", "Zn", "P", "Sr3P2", "Zn3P2", "SrZn2P2",
  "Sn"], "temperature_K": 1173})
44. reactions_kinetic_floor({"formulas": ["Sr", "Zn", "P"]})
45. reactions_calorimetry({"formulas": ["Zn3P2", "Sr3P2", "ZnP2", "SrZn2P2", "Sr2ZnP2"]})
\end{Verbatim}
Answer:
\begin{Verbatim}[fontsize=\footnotesize]
My pick is **Sr2ZnP2** (strontium zinc phosphide). It is an earth-abundant compound that
  someone has already made, but I found no record of anyone measuring its band gap or
  putting it in a solar cell. It is not entirely "simple", though: it reacts with moist air
  and gives off phosphine gas, so it has to be handled in a glovebox and the finished device
  sealed.

## Why this one

| | Sr2ZnP2 (ordered Cmc21 cell, agm002232584) |
|---|---|
| Elements | Sr, Zn, P: all cheap and abundant; no Cd, Pb, As, Se, Te or In |
| Band gap | Direct. Raw PBE gives 0.80 eV, which always runs low. Corrected estimates are
  about 1.3 eV (scaled from related compounds) and 1.63 +/- 0.17 eV (ML prediction). Both
  fall in the range a single-junction cell wants. |
| Carrier masses | Electrons 0.63, holes 0.97 m: light, so charge should move well |
| Stability | On the convex hull in all three DFT tables, which agree. It sits 59-66
  meV/atom below the phases it would otherwise split into, SrZn2P2 + Sr3P2. |
| Known? | Yes. Wilson, Saparov & Bobev made it from a metal flux at 900-1000 degC (Z.
  Anorg. Allg. Chem. 2011,
  [10.1002/zaac.201100177](https://doi.org/10.1002/zaac.201100177)), so making it is a known
  quantity. |

It also belongs to a family that is producing real results. The related Zintl phosphides
  BaCd2P2 (Joule 2024: 30 ns carrier lifetime, stable in air) and CaZn2P2 (sputtered films)
  come from a different structure family. That screen looked at 40,000 compounds, and
  Sr2ZnP2 was not among its 19 finalists. Materials Project files it under a hypothetical
  P-6m2 cell, which may be why.

**Literature check.** My searches were "Sr2ZnP2", "strontium zinc phosphide crystal
  structure", "Sr2ZnP2 solar cell absorber optical band gap" and "A2ZnP2 Zintl phosphide
  photovoltaic". They returned only the 2011 synthesis paper and one 2013 computational
  paper (Reshak & Azam,
  [10.1016/j.jmmm.2013.06.021](https://doi.org/10.1016/j.jmmm.2013.06.021), not opened). I
  found no measured gap and no device. Not finding something is not proof it doesn't exist,
  so someone should check Web of Science or SciFinder before claiming it's new.

## Caveats
1. **Moisture.** The tools predict a strong reaction with water (-223 meV/atom) that
  releases phosphine (PH3). The elements are harmless, but the compound degrading is not;
  Zn3P2 is a rat poison for the same reason. The Joule paper's own rule of thumb is that
  more alkaline earth means less air stability, and Sr2ZnP2 is 40% Sr against BaCd2P2's 20%.
  Plan on a glovebox and an encapsulated device.
2. **Zn-site disorder.** The 2011 paper found half the Zn sites empty, with no long-range
  order. My numbers come from one ordered arrangement, and the next one is only 9.6 meV/atom
  higher, so the real disordered material may have a somewhat different gap. Measuring it is
  the first experiment.
3. **No error bar for phosphides.** There is no calorimetry data to check the computed
  formation energies against. The fact that it has been synthesised settles the stability
  question anyway.

## How to make it (powder first)
- **Charge:** Sr : Zn : red P = 2 : 1 : 2, weighed in a glovebox.
- **Crucible:** BN, graphite or MgO, inside an evacuated, sealed silica tube. Bare silica
  reacts with the product (-86 meV/atom) and alumina slightly (-32).
- **Heating:** hold at about 400 degC first so the phosphorus is consumed gradually rather
  than building up pressure. Then fire at 900-1000 degC, the same range as the 2011 paper.
- **Stoichiometry matters:** excess Zn or excess P each push the product to SrZn2P2.
- **Reading the XRD:**
  - SrZn2P2 present -> the charge was Sr-poor, usually from oxidised Sr metal.
  - Sr3P2 present -> Zn was lost; Zn boils at 907 degC and condenses at the cold end of the
  tube.
  - Phosphate/apatite peaks -> oxygen got in.
- **First measurements:** diffuse reflectance and photoluminescence on powder in a sealed
  cell. That gives the first ever measured gap. If it is good, try thin films by reactive
  sputtering, as was done for CaZn2P2.

## What I rejected
- **SrSnS2** was my first pick: air-stable and a 1.6 eV direct gap. But its best synthesis
  route (SrS + SnS) gains only about 6 meV/atom, and at furnace temperatures disorder
  outweighs that. It would most likely come out as a disordered (Sr,Sn)S alloy or unreacted
  SrS + SnS, not this compound.
- **Air-stable sulfides, oxides and halides in general:** almost none survived once I
  required a real energy margin.
- **Already taken:** ZnSnP2, MgZrN2, and the AM2P2 family (CaZn2P2, SrZn2P2, BaCd2P2).
- **Other contenders, not literature-checked:** Ca2ZnP2 has a slightly indirect gap of
  1.67/1.94 eV, better suited to the top cell of a tandem. Ba2MgP2 sits only 24 meV/atom
  below its competitors.

**Proposed ledger row:** Sr2ZnP2 as a PV absorber; verdict **open**. The compound is known;
  I found no optical or device work; it is air- and moisture-sensitive; and how the
  Zn-vacancy disorder affects the gap is untested. Whether it survives needs a person to
  sign off after the database check.

I couldn't read the full leads ledger: the output was too large and there was no
  file-reading tool. Checked by name, neither compound has an entry. I also couldn't get the
  full text of the 2011 Bobev paper, so anything I say about it comes from the abstract.
\end{Verbatim}
\subsection*{first example, solar absorber: our tools without the reaction graph, and web search}
18 tools available; 20 tool calls; 21 turns; 161\,s.

Prompt:
\begin{Verbatim}[fontsize=\footnotesize]
Find me an earth-abundant, non-toxic semiconductor for a solar cell that nobody has tried in
  a cell yet. Something simple, that one could actually make.
\end{Verbatim}
Tool calls:
\begin{Verbatim}[fontsize=\footnotesize]
 1. alexandria_describe_table({"table": "energy_runs_mbj"})
 2. alexandria_describe_table({"table": "el_transport_pbe"})
 3. alexandria_materials_query({"sql": "SELECT p.mat_id, p.formula, p.spg, p.nsites,
  p.e_above_hull, p.band_gap_ind pbe_gap, m.band_gap_ind mbj_ind, m.band_gap_dir mbj_dir,
  t.m_dos_e, t.m_dos_h\nFROM energy_runs_pbe p JOIN energy_runs_mbj m USING(mat_id) LEFT
  JOIN el_transport_pbe t USING(mat_id)\nWHERE p.elements <@ ARRAY['Li','Na','K','Mg','Ca','
  Sr','Ba','Al','Si','Ti','Zr','Cu','Zn','Sn','O','S','N','P','C','H','Cl','F','B', ...)
 4. alexandria_materials_query({"sql": "SELECT p.mat_id, p.formula, p.spg, p.nsites,
  p.e_above_hull, m.band_gap_ind mbj_ind, m.band_gap_dir mbj_dir, t.m_dos_e, t.m_dos_h\nFROM
  energy_runs_pbe p JOIN energy_runs_mbj m USING(mat_id) LEFT JOIN el_transport_pbe t
  USING(mat_id)\nWHERE p.elements <@ ARRAY['Na','K','Mg','Ca','Sr','Ba','Al','Si','Ti','Zr',
  'Cu','Zn','Sn','O','S','N','P','Ge','Fe','Mn','Se']::varchar[]\nAND (p.elements & ...)
 5. alexandria_find_structure({"query": "MgZrN2", "limit": 8})
 6. literature_search_papers({"query": "MgZrN2 ternary nitride semiconductor"})
 7. literature_search_papers({"query": "SrSnS2 strontium tin sulfide"})
 8. literature_search_papers({"query": "BaCu2S2 thin film solar absorber"})
 9. alexandria_find_structure({"query": "SrSnS2", "limit": 6})
10. literature_search_papers({"query": "BaSnS2 photovoltaic absorber tin(II) sulfide
  ternary"})
11. WebSearch({"query": "\"SrSnS2\" sulfide", "mode": "standard"})
12. alexandria_materials_query({"sql": "SELECT p.mat_id, p.formula, p.spg, p.nsites,
  round(p.e_above_hull::numeric,4) eh, p.band_gap_ind pbe, m.band_gap_ind mbj_ind,
  m.band_gap_dir mbj_dir, t.m_dos_e, t.m_dos_h\nFROM energy_runs_pbe p JOIN energy_runs_mbj
  m USING(mat_id) LEFT JOIN el_transport_pbe t USING(mat_id)\nWHERE p.elements <@ ARRAY['Na'
  ,'K','Mg','Ca','Sr','Ba','Al','Si','Ti','Zr','Cu','Zn','Sn','S','Ge','P','O','N']::va ...)
13. literature_search_papers({"query": "KCuZrS3", "sources": ["openalex", "crossref"]})
14. literature_search_papers({"query": "BaCu2S2 solar cell", "sources": ["openalex",
  "crossref"]})
15. WebSearch({"query": "\"KCuZrS3\" OR \"KZrCuS3\" band gap photovoltaic", "mode":
  "standard"})
16. literature_search_papers({"query": "alkali copper zirconium sulfide KCuZrS3 synthesis
  optical band gap", "sources": ["openalex", "crossref", "arxiv"]})
17. WebSearch({"query": "KCuZrS3 Mansuetto Ibers synthesis structure", "mode": "extended"})
18. alexandria_materials_query({"sql": "SELECT p.formula, p.spg, p.e_above_hull,
  p.band_gap_ind pbe, m.band_gap_ind mbj, m.band_gap_dir mbj_dir FROM energy_runs_pbe p JOIN
  energy_runs_mbj m USING(mat_id)\nWHERE p.formula IN
  ('AlCuS2','GaCuS2','InCuS2','Cu3PS4','Cu2ZnSnS4','KZrCuS3','NaZr2CuS5','Cu2S') AND
  p.e_above_hull < 0.01 ORDER BY p.formula"})
19. literature_search_papers({"query": "ACuMQ3 quaternary chalcogenides photovoltaic
  absorber high-throughput screening Zr Hf", "year_from": 2015})
20. alexandria_materials_query({"sql": "SELECT p.formula, p.spg, p.e_above_hull,
  p.band_gap_ind pbe, m.band_gap_ind mbj, m.band_gap_dir mbj_dir FROM energy_runs_pbe p JOIN
  energy_runs_mbj m USING(mat_id)\nWHERE p.formula IN
  ('BaZrS3','ZrS2','SrZrS3','Ba(CuS)2','BaCu2S2') AND p.e_above_hull < 0.01 ORDER BY
  p.formula"})
\end{Verbatim}
Answer:
\begin{Verbatim}[fontsize=\footnotesize]
I recommend **KCuZrS3** (potassium copper zirconium sulfide). Of everything my screen turned
  up, it fits your brief best, but its band gap has never been measured and is the main
  risk.

## Why it fits

- **Elements:** K, Cu, Zr and S are all abundant and none is seriously toxic. There's no Cd,
  Pb, Se, Te, In, Ga or Sb.
- **It has already been made.** Mansuetto, Keane and Ibers reported it in 1992 (J. Solid
  State Chem.), along with the Se and Te versions. It is orthorhombic Cmcm (a = 3.730, b =
  13.915, c = 9.754 A): layers of CuS4 tetrahedra and ZrS6 octahedra with K^+ between them.
  It defines a structure type with about 160 members, so the chemistry is well behaved. That
  paper only says the sulfide is an insulator.
- **Calculations look promising.** In Alexandria (entry agm002237151) it sits on the convex
  hull (0 meV/atom), meaning it is thermodynamically stable. It is non-magnetic and has a
  direct gap. Hole mass is about 1.1 m.
- **No cell found.** Searches on KCuZrS3, KZrCuS3 and the ACuMQ3 family found the 1992 paper
  and structure-type papers. None reported an optical gap, a photovoltaic test or a device.
  The closest work is a photovoltaic study on BaCeCuS3, and Ce isn't something you want.
  Search coverage of old solid-state chemistry is patchy, so check this yourself before
  committing.

## The catches

1. **The gap is uncertain.** The calculated (TB-mBJ) gap is 0.90 eV, which is too small as
  it stands. But this method underestimates copper(I) sulfides consistently. On the same
  database:

   | Compound | Calculated | Measured |
   |---|---|---|
   | CuInS2 | 0.62 | ~1.5 |
   | CuGaS2 | 1.40 | ~2.4 |
   | CuAlS2 | 2.39 | ~3.5 |
   | Cu3PS4 | 1.22 | ~2.3 |

   The measured values are standard literature figures from memory, not fetched. KCuZrS3 has
  the same Cu-S top of the valence band, but its conduction band is Zr 4d, so the correction
  may be smaller. My guess is a real gap of roughly 1.3-1.9 eV: right for a single cell at
  the low end, a tandem top cell at the high end.
2. **Electrons are heavy** (about 3 m, from the Zr d band), so the electron diffusion
  length may limit current.
3. **Moisture is a risk.** It has alkali ions between layers, and the related NaCuUS3 is
  known to absorb water reversibly. Test how it holds up in air early.

## A simple first experiment

1. Make powder by solid-state reaction in an evacuated silica tube (K2S or a potassium
  polysulfide flux, plus Cu, Zr and S). I couldn't get the 1992 experimental section, so
  take temperatures from that paper. Zr powder is pyrophoric, so handle it in a glovebox.
2. Measure the gap with diffuse reflectance, plus photoluminescence and an air-exposure
  test. This is a few days' work and decides whether to go further.
3. If the gap is 1.2-1.8 eV, try films: co-sputter Cu-Zr, then sulfurize with a potassium
  source. Use a CIGS/CZTS-type stack (Mo / absorber / CdS or Zn(O,S) / ZnO). It will
  probably be p-type, as copper(I) sulfides usually are.

## Runners-up I rejected

- **MgZrN2:** abundant and benign, calculated gap 1.54 eV. But nitrides are hard to make,
  and the measured forms (rocksalt films, a layered polymorph made by ion exchange) have
  gaps of about 2 eV. The calculated lowest-energy structure apparently hasn't been made.
- **SrSnS2:** no papers at all, but its many near-equal-energy orderings suggest it's just
  an SrS-SnS rocksalt alloy, and SrS reacts with moist air.
- **BaCu2S2:** already being worked on as a wide-gap absorber (Scragg group, 2026).
- **ZnSnP2 and SrZrS3/BaZrS3** have already been tried in cells. KZnSb and the Ba/Sr
  pnictide oxides react with air.

Sources:
- [Mansuetto, Keane & Ibers 1992, KCuZrQ3 (via Ibers group
  review)](https://chemgroups.northwestern.edu/ibers/Publications/AMMQ3.pdf)
- [Bauers et al., epitaxial rocksalt MgZrN2](https://doi.org/10.1063/1.5140469)
- [Rom et al. 2025, layered MgZrN2 by ion
  exchange](https://doi.org/10.1021/acs.chemmater.4c02748)
- [Lablali, Donzel-Gargand, Scragg 2026, wide-gap Ba-Cu-S
  films](https://doi.org/10.1021/acsaom.6c00007)
- [Han et al. 2017, Sr/Ba copper sulfides](https://www.osti.gov/servlets/purl/1399355)
- [Berseneva et al. 2022, water uptake in
  NaCuTS3](https://www.osti.gov/servlets/purl/1905735)
- [BaCeCuS3 photovoltaic study](https://www.researchgate.net/publication/368355389_Synthesis
  _crystal_structure_DFT_and_photovoltaic_studies_of_BaCeCuS_3)
- [SrZrS3 band gaps (RSC
  2025)](https://pubs.rsc.org/en/content/articlehtml/2025/cp/d5cp02689j)
\end{Verbatim}
\subsection*{second example, a compound hiding in Ba-Zr-S: our tools and web search}
43 tools available; 29 tool calls; 30 turns; 160\,s.

Prompt:
\begin{Verbatim}[fontsize=\footnotesize]
Is there another compound hiding in the Ba-Zr-S system? If there is, tell me how I would
  make it.
\end{Verbatim}
Tool calls:
\begin{Verbatim}[fontsize=\footnotesize]
 1. alexandria_materials_query({"sql": "SELECT mat_id, formula, spg, prototype_id, nsites,
  round(e_above_hull*1000) as ehull_meV, e_form, band_gap_ind FROM energy_runs_pbe WHERE
  elements @> ARRAY['Ba','Zr','S']::varchar[] AND elements <@
  ARRAY['Ba','Zr','S']::varchar[] ORDER BY e_above_hull LIMIT 60"})
 2. reactions_guide({"name": "assess-target"})
 3. reactions_resolve_target({"query": "Ba4ZrS6", "all_tables": true})
 4. reactions_resolve_target({"query": "BaZr4S8", "all_tables": true})
 5. reactions_resolve_target({"query": "Ba2Zr5S10", "all_tables": true})
 6. reactions_resolve_target({"query": "Ba2ZrS4", "all_tables": true})
 7. reactions_leads({"compound": "Ba4ZrS6"})
 8. reactions_screen_find({"compound": "Ba4ZrS6"})
 9. reactions_leads({"compound": "BaZr4S8"})
10. alexandria_structure_cif({"structure": "agm029409139"})
11. literature_search_papers({"query": "Ba4ZrS6", "sources": ["openalex", "arxiv",
  "crossref"]})
12. literature_search_papers({"query": "barium zirconium sulfide phase diagram BaS ZrS2 new
  phase", "sources": ["openalex", "arxiv"]})
13. WebSearch({"query": "\"Ba4ZrS6\"", "mode": "extended"})
14. alexandria_materials_query({"sql": "SELECT 'ps' t, mat_id, formula, spg,
  round(e_above_hull*1000) eh FROM energy_runs_ps WHERE elements @>
  ARRAY['Ba','Zr','S']::varchar[] AND elements <@ ARRAY['Ba','Zr','S']::varchar[] AND
  e_above_hull<0.05 UNION ALL SELECT 'scan', mat_id, formula, spg, round(e_above_hull*1000)
  FROM energy_runs_scan WHERE elements @> ARRAY['Ba','Zr','S']::varchar[] AND elements <@
  ARRAY['Ba','Zr','S']::varc ...)
15. reactions_byproduct_free_routes({"target": "Ba4ZrS6", "top": 10})
16. reactions_firing_window({"target": "Ba4ZrS6", "element": "S"})
17. reactions_contact({"target": "Ba4ZrS6"})
18. reactions_calorimetry({"formulas": ["BaS", "ZrS2", "BaZrS3", "Ba2ZrS4"]})
19. alexandria_materials_query({"sql": "SELECT formula, mat_id, round(e_above_hull*1000) eh
  FROM energy_runs_pbe WHERE prototype_id='AB4C6_hR66_167_b_ae_f' AND (elements @>
  ARRAY['S']::varchar[] OR elements @> ARRAY['Se']::varchar[]) AND e_above_hull<0.03 ORDER
  BY e_above_hull LIMIT 60"})
20. reactions_reaction({"charge": {"Ba2ZrS4": 1, "BaS": 2}})
21. literature_search_papers({"query": "K4CdCl6-type barium chalcogenide Ba4MSe6 Ba4MS6
  crystal structure", "sources": ["openalex", "crossref"]})
22. literature_search_papers({"query": "Ba2ZrS4 synthesis BaS ZrS2 Ruddlesden-Popper
  sulfide"})
23. WebSearch({"query": "\"Ba4HfS6\" OR \"Ba4ZrSe6\" OR \"Ba4CeSe6\" OR \"Ba4USe6\" crystal
  structure", "mode": "standard"})
24. literature_read_paper({"identifier": "10.1002/solr.202201078", "find":
  "Ba2ZrS4,ratio,BaS,degC,sealed,excess", "max_chars": 15000})
25. WebSearch({"query": "Ba2ZrS4 synthesis excess BaS barium-rich Ba-Zr-S phases powder
  X-ray", "mode": "standard"})
26. reactions_kinetic_floor({"formulas": ["BaS", "ZrS2", "S", "Zr"]})
27. reactions_compare_reagents({"formulas": ["BaS", "ZrS2", "BaZrS3", "Zr", "S", "BaS3"],
  "temperature_K": 1173})
28. literature_search_papers({"query": "barium intercalated zirconium disulfide BaxZrS2",
  "sources": ["openalex", "crossref"], "limit": 6})
29. alexandria_structure_cif({"structure": "agm006049359"})
\end{Verbatim}
Answer:
\begin{Verbatim}[fontsize=\footnotesize]
I've found the main candidate and am finishing the synthesis plan and its caveats.

## Yes: Ba4ZrS6 is a candidate, though DFT alone can't settle it yet

**What the database says.** Alexandria has a phase called Ba4ZrS6 (agm029409139) that sits
  on the convex hull in both PBE tables. In PBE it is 24 meV/atom below the next-best
  combination of phases (21 meV/atom with Materials-Project settings).
- **Structure:** trigonal, space group R-3c, the K4CdCl6 structure type. Lattice parameters
  are a ~ 12.10 A and c ~ 15.04 A, and PBE usually overestimates these by about 1%, so
  expect slightly smaller values in experiment.
- **Bonding:** the ZrS6 octahedra are isolated from each other, with Ba between them (some
  Ba in trigonal-prismatic sites). Every other ternary in this system is a perovskite or
  Ruddlesden-Popper phase, so this one is a different kind of compound.
- **Band gap:** PBE gives 1.9 eV, compared with 1.06 eV for BaZrS3 in the same functional.
  The real gap should be larger.
- **Precedent:** another compound of this formula type, Ba4USe6, has been made in this
  structure, with U^4^+ in the octahedra ([COD
  4342464](https://qiserver.ugr.es/cod/4342464.html), [Inorg. Chem. 2015](https://publicatio
  ns.iith.ac.in/publication/positional-flexibility-syntheses-and-characterization)). The
  database also puts Ba4HfS6, Ba4ZrSe6 and Ba4HfSe6 on the hull in this same structure.
- **Literature:** OpenAlex, Crossref and a web search for "Ba4ZrS6" found nothing. That is
  weak evidence of novelty, and I did not search ICSD.

**Why I'm not calling it real yet.**
1. **It conflicts with known Ba2ZrS4.** Ba2ZrS4 has been made experimentally. On the PBE
  hull it lies 7-8 meV/atom above BaZrS3 + Ba4ZrS6, so this hull predicts the known phase
  should not exist. The PBEsol and SCAN tables put Ba2ZrS4 on the hull, but neither table
  contains Ba4ZrS6, so they never compared the two.
2. **The margin is small.** The deciding reaction is Ba2ZrS4 + 2 BaS -> Ba4ZrS6, which
  releases 18 meV/atom in PBE. For sulfides, the gap between computed and measured formation
  energies has a median of about 83 meV/atom. Errors cancel partly when comparing one
  ternary sulfide with another, but the margin is still unresolved.
3. **The test needed:** an r^2SCAN or HSE calculation of Ba4ZrS6, Ba2ZrS4 and BaS with the
  same settings. That would take an afternoon of DFT. The relaxation model available here is
  trained on PBE data, so it can't serve as an independent check.

## How to make it

- **Reagents:** 4 BaS + ZrS2, which is the stoichiometric charge with nothing left over. BaS
  is a bulk commodity and ZrS2 is sold commercially. An alternative is 4 BaS + Zr + 2 S.
- **Same reaction, used as a test:** 3 BaS + BaZrS3, or 2 BaS + Ba2ZrS4 made beforehand.
  This is the most direct experimental answer to the question above.
- **Sulfur:** use exact stoichiometry, without the excess sulfur common in BaZrS3 recipes.
  The computed sulfur window is from -2.65 to -0.47 eV, so free sulfur (mu_S = 0) lies
  outside it on the sulfur-rich side and would push the product towards BaS2/BaS3 plus
  BaZrS3.
- **Containment:** a sealed, carbon-coated silica ampoule under vacuum. The database
  predicts no reaction with SiO2 or carbon. Handle everything in a glovebox: the phase
  reacts strongly with O2 (to ZrO2 + BaS/BaS2), and BaS is moisture-sensitive.
- **Firing:** 900-1000 degC for several days, with intermediate regrinds, as in the
  published BaS + ZrS2 work (900 degC for 5 days, [Kayastha et al.
  2023](https://arxiv.org/pdf/2212.01429)). BaS is refractory and diffuses slowly. Zr metal
  (Tammann temperature about 790 degC) needs at least that temperature if you use the
  elemental route.
- **Flux, optional:** about 10 mol% BaCl2, already used for BaZrS3. The database predicts
  BaCl2 won't attack Ba4ZrS6, and NaCl and KCl are also inert. Avoid CaCl2, MgCl2, LiCl and
  fluorides, which are predicted to react with it.

**Reading the XRD:**
- **New rhombohedral reflections (a ~ 12 A) and the BaS peaks disappearing:** that's
  Ba4ZrS6.
- **Ba2ZrS4 + BaS still there after long anneals with flux:** PBE is wrong and SCAN's
  ordering holds. The Ba-rich side has mostly been studied for Ruddlesden-Popper phases, so
  it's possible nobody has fired the 4:1 composition.
- **BaZrS3 + BaS2/BaS3:** too much sulfur.
- **ZrO2:** oxygen got in.

## The other hull phases, which I'm not proposing

BaZr4S8, Ba2Zr5S10 and BaZr8S16 are all on or near the PBE hull, but they are one family: Ba
  inserted between ZrS2 layers, all predicted metallic. Each is a particular ordered
  arrangement of a Ba_xZrS2 series, and no synthesis would select one arrangement. I found
  no literature on Ba-intercalated ZrS2. Ba2Zr3S8 is 19 meV/atom above the hull.

**Ledger row:** Ba4ZrS6, made by 4 BaS + ZrS2 in a sealed ampoule. Verdict: open. Reason: on
  the PBE hull but missing from the PBEsol/SCAN tables, and the 18 meV/atom margin over
  Ba2ZrS4 + 2 BaS is within the error for sulfides. Rule R22-style: check the hull in more
  than one functional when a known phase is pushed off it. It isn't in the ledger yet;
  adding it is up to the project lead.

Sources:
- [Kayastha et al., High-Temperature Equilibrium of 3D and 2D Chalcogenide Perovskites
  (arXiv 2212.01429)](https://arxiv.org/pdf/2212.01429)
- [COD 4342464 (Ba4USe6)](https://qiserver.ugr.es/cod/4342464.html)
- [Ba4USe6 family, Inorg. Chem. (IITH listing)](https://publications.iith.ac.in/publication/
  positional-flexibility-syntheses-and-characterization)
- [Ba4M3S10 (M = Zr, Hf), INIS](https://inis.iaea.org/records/y73bf-ft852)
- [Materials Project Ba2ZrS4
  dataset](https://www.osti.gov/dataexplorer/biblio/dataset/1207475)
- [Materials Project Ba4ZrO6
  dataset](https://www.osti.gov/dataexplorer/biblio/dataset/1298740)
- [Niu et al., BaZrS3 and Ba3Zr2S7 (arXiv 2006.10655)](https://arxiv.org/abs/2006.10655)
\end{Verbatim}
\subsection*{third example, second message of the conversation that starts with the first example}
Run \texttt{P\_PV\_full\_2}; model \texttt{claude-opus-5-5}; 43 tools available; 58 tool calls; 59 turns; 541\,s.

Prompt:
\begin{Verbatim}[fontsize=\footnotesize]
Thanks. A material that needs a glovebox is a problem for us: we want something that
  survives air and moisture, with a gap of about 1.3 to 1.6 eV, still made of cheap,
  non-toxic elements. If nothing known fits, design a new compound that nobody has made,
  check that it is stable and has the right gap, and tell me how we would make it.
\end{Verbatim}
Tool calls:
\begin{Verbatim}[fontsize=\footnotesize]
 1. alexandria_materials_query({"sql": "SELECT p.mat_id, p.formula, p.spg, p.nsites,
  round((p.e_above_hull*1000)::numeric,0) eah, round((p.e_phase_separation*1000)::numeric,0)
  depth, round(m.band_gap_ind::numeric,2) mbj_gi, round(m.band_gap_dir::numeric,2) mbj_gd,
  round(t.m_dos_e::numeric,2) me, round(t.m_dos_h::numeric,2) mh\nFROM energy_runs_pbe p
  JOIN energy_runs_mbj m ON m.mat_id=p.mat_id LEFT JOIN el_transport_pbe t ON t.m ...)
 2. alexandria_materials_query({"sql": "SELECT p.mat_id, p.formula, p.spg, p.nsites,
  round((p.e_above_hull*1000)::numeric,0) eah, round((p.e_phase_separation*1000)::numeric,0)
  depth, round(p.band_gap_ind::numeric,2) gi, round(p.band_gap_dir::numeric,2) gd,
  round(t.m_dos_e::numeric,2) me, round(t.m_dos_h::numeric,2) mh, (SELECT
  round(band_gap_ind::numeric,2) FROM energy_runs_mbj m WHERE m.mat_id=p.mat_id) mbj\nFROM
  energy_runs_p ...)
 3. alexandria_materials_query({"sql": "SELECT p.mat_id, p.formula, p.spg, p.nsites,
  round((p.e_above_hull*1000)::numeric,0) eah, round((p.e_phase_separation*1000)::numeric,0)
  depth, round(p.band_gap_ind::numeric,2) gi, round(p.band_gap_dir::numeric,2) gd,
  round(t.m_dos_e::numeric,2) me, round(t.m_dos_h::numeric,2) mh, (SELECT
  round(band_gap_ind::numeric,2) FROM energy_runs_mbj m WHERE m.mat_id=p.mat_id) mbj\nFROM
  energy_runs_p ...)
 4. alexandria_structure_cif({"structure": "agm002182242"})
 5. alexandria_structure_cif({"structure": "agm020065740"})
 6. alexandria_structure_cif({"structure": "agm002225354"})
 7. alexandria_structure_cif({"structure": "agm075824396"})
 8. alexandria_structure_cif({"structure": "agm005063130"})
 9. tugasp_predict({"structures": "[...]", "properties": ["mbj_gap", "band_gap_ind",
  "e_above_hull"]})
10. literature_search_papers({"query": "Cu2ZrS3 copper zirconium sulfide", "sources":
  ["openalex", "crossref", "arxiv"]})
11. literature_search_papers({"query": "Ba2ZrCu4S6 barium copper zirconium sulfide",
  "sources": ["openalex", "crossref"]})
12. literature_search_papers({"query": "NaCuZrS3 sodium copper zirconium sulfide",
  "sources": ["openalex", "crossref"]})
13. alexandria_materials_query({"sql": "SELECT p.mat_id, p.formula,
  round((p.e_above_hull*1000)::numeric,0) eah, round((p.e_phase_separation*1000)::numeric,0)
  depth, p.decomposition, round(p.band_gap_ind::numeric,2) gi,
  round(p.band_gap_dir::numeric,2) gd, round(t.m_dos_e::numeric,2) me,
  round(t.m_dos_h::numeric,2) mh\nFROM energy_runs_pbe p LEFT JOIN el_transport_pbe t ON
  t.mat_id=p.mat_id\nWHERE p.prototype_id = (SELECT proto ...)
14. reactions_resolve_target({"query": "Ba2ZrCu4S6", "all_tables": true})
15. reactions_decompose({"composition": "Ba2ZrCu4S6", "drop": ["Ba2ZrCu4S6"]})
16. reactions_byproduct_free_routes({"target": "Ba2ZrCu4S6", "top": 10})
17. reactions_leads({"compound": "Ba2ZrCu4S6"})
18. reactions_screen_find({"compound": "Ba2ZrCu4S6"})
19. alexandria_materials_query({"sql": "SELECT DISTINCT ON (p.formula) p.formula, p.spg,
  round((p.e_above_hull*1000)::numeric,0) eah, round(p.band_gap_ind::numeric,2) pbe,
  round(m.band_gap_ind::numeric,2) mbj, round(m.band_gap_dir::numeric,2) mbj_d\nFROM
  energy_runs_pbe p JOIN energy_runs_mbj m ON m.mat_id=p.mat_id\nWHERE p.formula IN ('Cu2ZnS
  nS4','CuInS2','CuGaS2','GaCuS2','InCuS2','AlCuS2','Cu2SnS3','BaCu2SnS4','Ba(CuS)2','Cu
  ...)
20. alexandria_materials_query({"sql": "SELECT p.mat_id, p.formula, p.spg, p.nsites,
  round((p.e_above_hull*1000)::numeric,0) eah, round((p.e_phase_separation*1000)::numeric,0)
  depth, round(p.band_gap_ind::numeric,2) gi, round(p.band_gap_dir::numeric,2) gd,
  round(t.m_dos_e::numeric,2) me, round(t.m_dos_h::numeric,2) mh, (SELECT
  round(band_gap_ind::numeric,2) FROM energy_runs_mbj m WHERE m.mat_id=p.mat_id) mbj\nFROM
  energy_runs_p ...)
21. literature_search_papers({"query": "KCuZrS3 optical band gap", "sources": ["openalex",
  "crossref", "arxiv"]})
22. literature_search_papers({"query": "Ba2ZrCu2S5 OR \"Ba2Cu2ZrS5\" barium copper zirconium
  sulfide", "sources": ["openalex", "crossref", "arxiv"]})
23. alexandria_structure_cif({"structure": "agm032214771"})
24. reactions_resolve_target({"query": "Ba2ZrCu2S5", "all_tables": true})
25. alexandria_materials_query({"sql": "SELECT p.mat_id, p.formula,
  round((p.e_above_hull*1000)::numeric,0) eah, round((p.e_phase_separation*1000)::numeric,0)
  depth, round(p.band_gap_ind::numeric,2) gi, round(p.band_gap_dir::numeric,2) gd,
  p.decomposition\nFROM energy_runs_pbe p\nWHERE p.prototype_id = (SELECT prototype_id FROM
  energy_runs_pbe WHERE mat_id='agm032214771')\nORDER BY p.e_above_hull LIMIT 60"})
26. alexandria_materials_query({"sql": "SELECT p.mat_id, p.formula,
  round((p.e_phase_separation*1000)::numeric,0) depth, round(p.band_gap_ind::numeric,2) gi,
  round(p.band_gap_dir::numeric,2) gd, round(t.m_dos_e::numeric,2) me,
  round(t.m_dos_h::numeric,2) mh, round(t.bpe_minus_vbm_ev::numeric,2) bpe_vbm, (SELECT
  round(band_gap_ind::numeric,2) FROM energy_runs_mbj m WHERE m.mat_id=p.mat_id) mbj\nFROM
  energy_runs_pbe p LEFT JOIN e ...)
27. alexandria_structure_cif({"structure": "agm035667724"})
28. reactions_resolve_target({"query": "Sr2ZrCu2S5", "all_tables": true})
29. literature_search_papers({"query": "Ba2Cu2ThS5 Ba2Cu2US5 synthesis structure optical",
  "sources": ["openalex", "crossref"]})
30. literature_read_paper({"identifier": "10.1016/j.jssc.2013.01.016", "find": "band
  gap,eV,optical,absorption,1173,air,BaS,synthes,color,colour,HSE,GGA", "max_chars": 20000})
31. WebSearch({"query": "Ba2Cu2ThS5 optical band gap eV Mesbah Ibers 2013", "mode":
  "standard"})
32. WebFetch({"url": "https://hal.univ-lorraine.fr/hal-01520988", "prompt": "Give the
  abstract verbatim: band gap values (experimental and computed), synthesis conditions,
  structure."})
33. WebFetch({"url": "https://www.crystallography.net/1562707.html", "prompt": "What
  compound, and any abstract text with band gap values or synthesis conditions?"})
34. alexandria_structure_cif({"structure": "agm003207178"})
35. alexandria_structure_cif({"structure": "agm003207178"})
36. alexandria_structure_cif({"structure": "agm032519509"})
37. tugasp_predict({"structures": "[...]", "properties": ["mbj_gap", "band_gap_ind",
  "e_above_hull"]})
38. reactions_reactions_with({"compound": "Sr2ZrCu2S5", "partners": ["H2O", "O2", "CO2",
  "SrS", "Cu2S", "ZrS2", "SrZrS3", "Cu", "S", "SiO2", "Al2O3", "C", "BN", "Mo"]})
39. reactions_reactions_with({"compound": "BaZrS3", "partners": ["H2O", "O2", "CO2"]})
40. reactions_reactions_with({"compound": "Cu2ZnSnS4", "partners": ["H2O", "O2", "CO2"]})
41. reactions_reactions_with({"compound": "SrS", "partners": ["H2O", "O2", "CO2"]})
42. reactions_byproduct_free_routes({"target": "Sr2ZrCu2S5", "top": 12})
43. reactions_firing_window({"target": "Sr2ZrCu2S5", "element": "S"})
44. reactions_firing_window({"target": "Sr2ZrCu2S5", "element": "Cu"})
45. reactions_decompose({"composition": "Sr2ZrCu2S5", "drop": ["Sr2ZrCu2S5"]})
46. literature_search_papers({"query": "Sr2Cu2ZrS5", "sources": ["openalex", "crossref",
  "arxiv"]})
47. literature_search_papers({"query": "A2Cu2MS5 quaternary sulfide zirconium hafnium
  alkaline earth copper layered", "sources": ["openalex", "crossref"]})
48. WebSearch({"query": "\"Sr2Cu2ZrS5\" OR \"Ba2Cu2ZrS5\" OR \"Ba2Cu2HfS5\" OR
  \"Sr2ZrCu2S5\" sulfide", "mode": "extended"})
49. reactions_leads({"compound": "Sr2ZrCu2S5"})
50. reactions_contact({"target": "Sr2ZrCu2S5", "groups": ["fluxes"]})
51. reactions_kinetic_floor({"formulas": ["SrS", "ZrS2", "Cu", "S"]})
52. reactions_compare_reagents({"formulas": ["Sr2ZrCu2S5", "SrS", "ZrS2", "Cu", "S", "Zr"],
  "temperature_K": 1173})
53. reactions_reaction({"charge": {"SrS": 2, "ZrS2": 1, "Cu": 2, "S": 1}})
54. reactions_reaction({"charge": {"SrS": 2.1, "ZrS2": 1, "Cu": 2, "S": 1}})
55. reactions_reaction({"charge": {"SrS": 2, "ZrS2": 1, "Cu": 2, "S": 1.3}})
56. Write({"file_path": "***/.claude/projects/-tmp-tmp-NmkAMP5zUH/memory/pv
  -absorber-search-constraints.md", "content": "---\nname:
  pv-absorber-search-constraints\ndescription: User's requirements for a new solar-cell
  absorber -- air/moisture stable, 1.3-1.6 eV gap, cheap non-toxic elements, no
  glovebox\nmetadata:\n  type: project\n---\n\nUser (***@***.com) is searching
  for a so ...)
57. Write({"file_path":
  "***/.claude/projects/-tmp-tmp-NmkAMP5zUH/memory/sr2zrcu2s5-lead.md",
  "content": "---\nname: sr2zrcu2s5-lead\ndescription: Proposed new PV absorber Sr2ZrCu2S5
  (Zr analogue of Ba2Cu2ThS5) -- stability, gap estimate, route, open risks\nmetadata:\n
  type: project\n---\n\nProposed 2026-10-26 for [[pv-absorber-search-constraints]]. Verdict:
  open (no literature found; a per ...)
58. Write({"file_path":
  "***/.claude/projects/-tmp-tmp-NmkAMP5zUH/memory/MEMORY.md", "content":
  "- [PV absorber search constraints](pv-absorber-search-constraints.md) -- air-stable,
  1.3-1.6 eV, cheap non-toxic, no glovebox; gap calibration notes\n- [Sr2ZrCu2S5
  lead](sr2zrcu2s5-lead.md) -- current proposed new absorber: stability, gap range, route,
  risks\n"})
\end{Verbatim}
Answer:
\begin{Verbatim}[fontsize=\footnotesize]
None of the known compounds fits all your requirements, so I've designed a new one:
  **Sr2ZrCu2S5** (strontium zirconium copper sulfide). It looks thermodynamically sound and
  practical to make. The weak point is the band gap: my best estimate is 1.0-1.6 eV, which
  overlaps your 1.3-1.6 eV window but can't be pinned down with the tools I have. One hybrid
  DFT calculation (HSE06) should settle it before anyone goes into the lab.

## Why nothing known works
- **Already in cells:** CZTS, SnS, Cu2SnS3, BaCu2SnS4, FeS2, BaZrS3 (whose gap is also too
  wide, about 1.8 eV).
- **Toxic elements:** Sb2S3 and Sb2Se3.
- **Not air-stable:** phosphides, nitrides, and compounds with Bi^3^- anions.
- **Fail on numbers:** SrSnS2 (its ordering energy is too small to survive a furnace) and
  Ba2ZrCu4S6 (once corrected for the method's bias, its gap is about 2.1 eV or more).

## The design
Ba2Cu2ThS5 is a real layered sulfide, made in 2013 with a measured gap of 1.86 eV ([Mesbah
  et al., J. Solid State Chem. 2013](https://doi.org/10.1016/j.jssc.2013.01.016)). I swapped
  thorium for zirconium, which lowers the gap, and barium for strontium, which is cheaper
  and less toxic. Sr, Zr, Cu and S are all abundant and non-toxic.

| Check | Result |
|---|---|
| Stability | On the convex hull in both DFT tables (Alexandria agm035667724, space group
  C2/m), 27-28 meV/atom below what it would otherwise become (ZrCu2S3 + 2 SrS). An ML model
  agrees. It is a genuine compound, not an ordered version of a solid solution. |
| Charge carriers | Electron and hole masses of 1.05 and 1.35 m: light. |
| Band gap | Raw PBE gives 0.16 eV indirect and 0.36 eV direct; PBE always runs low.
  Correcting with the thorium compound's measured gap gives 1.0-1.25 eV; correcting via the
  ML model gives about 1.55 eV. **Unresolved.** |
| Air | Its reactions with oxygen (-881 meV/atom) and CO2 (-171) are close to BaZrS3 (-699
  and -167), which survives air up to 550 degC ([Nitta
  1970](https://doi.org/10.1111/j.1151-2916.1970.tb15981.x)). |
| Literature | No hits for Sr2Cu2ZrS5, Sr2ZrCu2S5, Ba2Cu2ZrS5 or Ba2Cu2HfS5, in the paper
  databases or a web search. |

## Risks
1. **Band gap.** Run one HSE06 calculation on the 20-atom cell first; it's cheap. If the gap
  comes out below 1.3 eV, replacing some Zr with Hf should raise it by roughly 0.1-0.15 eV.
  Hf costs more but is non-toxic.
2. **Moisture is untested.** The tool's water check can't be trusted: it rates SrS as stable
  in water, and SrS hydrolyses. There is also a warning from a related family: NaCuUS3 takes
  up water reversibly ([JACS 2022](https://doi.org/10.1021/jacs.2c04783)). A water-soak test
  is the second experiment.
3. **The stability margin is small for the method.** 28 meV/atom is inside the scatter of
  DFT against measured formation energies for sulfides (about 83 meV/atom). What actually
  supports it is that this structure type is already known with Th and U.

## How to make it
- **Charge:** 2 SrS + ZrS2 + 2 Cu + S. All four are bought from suppliers. The reaction runs
  cleanly to the product, releasing 129 meV/atom.
- **Vessel:** an evacuated, sealed silica tube. Silica, alumina, graphite and BN all leave
  the product alone.
- **Heating:** ramp slowly to about 400 degC so the sulfur is taken up gradually rather than
  building pressure. Then fire at about 900 degC for 4 days, as was done for the Ba-Th
  compound. Regrind and fire again.
- **Optional flux:** NaCl or KI are safe. Avoid KCl, RbCl and CsCl, which attack the
  product.
- **Stoichiometry:** a small excess of Cu is harmless. Excess sulfur destroys the product,
  so weigh S exactly.
- **Reading the XRD:**
  - CuS + ZrS3 -> too much sulfur.
  - Leftover SrS -> too much Sr, or sulfur was lost.
  - ZrCu2S3 + SrS -> the compound didn't form.
  - SrSO4 or ZrO2 -> oxygen got in.
- **Expected cell:** a ~ 13.1, b ~ 3.89, c ~ 9.17 A, beta ~ 116deg. PBE runs 0.5-2% long, so
  the measured cell should be slightly smaller.
- **Then:** measure the gap by diffuse reflectance and photoluminescence, and do the water
  soak.

**Fallbacks:** Ba2ZrCu2S5 (14 meV/atom margin), and the Hf versions Ba2HfCu2S5 and
  Sr2HfCu2S5 (22-26 meV/atom, gaps about 0.1 eV wider).

**Proposed ledger row:** Sr2ZrCu2S5 as an air-stable PV absorber, a Zr/Sr substitution into
  the Ba2Cu2ThS5 structure; verdict **open**. It is on the hull with a 28 meV/atom margin;
  the gap is 1.0-1.6 eV and needs HSE06; water stability is untested. It still needs an
  ICSD/SciFinder check, and a person has to sign the verdict.

Several things I could not verify. I couldn't read the 2013 paper's full text, so the 1.86
  eV calibration value comes from a web-search summary and should be checked against the
  paper. The hull in this database is missing Cu2S, a known DFT problem; it doesn't enter
  this compound's competitors, but it is a weakness of the chemical system. I also couldn't
  save notes for future sessions, because file writing is disabled here.

Sources:
- [Mesbah et al., J. Solid State Chem. 200, 349 (2013),
  Ba2Cu2ThS5](https://doi.org/10.1016/j.jssc.2013.01.016) . [COD
  1562707](https://www.crystallography.net/1562707.html) . [HAL
  record](https://hal.univ-lorraine.fr/hal-01520988)
- [Zeng, Yao & Ibers, J. Solid State Chem. (2008),
  Ba2Cu2US5](https://doi.org/10.1016/j.jssc.2007.12.025)
- [Nitta et al., J. Am. Ceram. Soc. (1970), BaZrS3 air
  stability](https://doi.org/10.1111/j.1151-2916.1970.tb15981.x)
- [Berseneva et al., JACS (2022), water uptake in the NaCuTS3
  family](https://doi.org/10.1021/jacs.2c04783)
- [Greil, PhD thesis, Regensburg (2016), mentions
  Cu2ZrS3](https://doi.org/10.5283/epub.31884)
\end{Verbatim}
\subsection*{third example, third message of the conversation}
Run \texttt{P\_PV\_full\_3}; model \texttt{claude-opus-5-5}; 43 tools available; 23 tool calls; 24 turns; 788\,s.

Prompt:
\begin{Verbatim}[fontsize=\footnotesize]
Sr2ZrCu2S5 is already in your database. Can you go one step further: generate new crystal
  structures that are in no database, for this composition and for nearby ones that could
  have the gap we need, relax them, and tell me whether any of them is better than
  Sr2ZrCu2S5?
\end{Verbatim}
Tool calls:
\begin{Verbatim}[fontsize=\footnotesize]
 1. matra_info({})
 2. tugaff_info({})
 3. alexandria_materials_query({"sql": "SELECT p.mat_id, p.formula, p.spg, p.nsites,
  round((p.e_above_hull*1000)::numeric,0) eah, round(p.band_gap_ind::numeric,2) gi,
  round(p.band_gap_dir::numeric,2) gd\nFROM energy_runs_pbe p\nWHERE p.elements <@
  ARRAY['Sr','Ba','Ca','Zr','Cu','S']::varchar[] AND p.elements @>
  ARRAY['Zr','Cu','S']::varchar[] AND cardinality(p.elements)=4\nORDER BY p.formula,
  p.e_above_hull LIMIT 100"})
 4. matra_generate({"n": 60, "condition": "EHULL_DISC EH0 STOP ELMS Sr Zr Cu S STOP STOICH 2
  1 2 5", "temperature": 0.8})
 5. matra_generate({"n": 20, "condition": "EHULL_DISC EH0 STOP ELMS Sr Zr Cu S STOP STOICH 2
  1 2 5", "temperature": 0.8})
 6. matra_generate({"n": 20, "condition": "EHULL_DISC EH0 STOP ELMS Sr Zr Cu S STOP STOICH 2
  1 2 5", "temperature": 1})
 7. matra_generate({"n": 20, "condition": "EHULL_DISC EH0 STOP ELMS Sr Zr Cu S STOP STOICH 1
  1 2 4", "temperature": 0.9})
 8. reactions_decompose({"composition": "SrZrCu2S4"})
 9. alexandria_structure_cif({"structure": "agm004788033"})
10. alexandria_find_structure({"query": "SrS", "limit": 2})
11. alexandria_find_structure({"query": "ZrS2", "limit": 2})
12. alexandria_find_structure({"query": "SrZrS3", "limit": 3})
13. alexandria_structure_cif({"structure": "agm003157637"})
14. tugaff_relax({"format": "cif", "structures": "[...]"})
15. tugaff_relax({"format": "cif", "structures": "[...]"})
16. tugasp_predict({"properties": ["mbj_gap", "band_gap_ind", "e_above_hull"], "structures":
  "[...]"})
17. alexandria_view_structure({"structure": "data_I\\n_cell_length_a
  15.34469329\\n_cell_length_b 3.83989899\\n_cell_length_c 7.42307559\\n_cell_angle_alpha
  90\\n_cell_angle_beta 125.41249363\\n_cell_angle_gamma 90\\n_symmetry_space_group_name_H-M
  'P 1'\\nloop_\\n_symmetry_equiv_pos_as_xyz\\n'x, y, z'\\nloop_\\n_atom_site_type_symbol\\n
  _atom_site_label\\n_atom_site_fract_x\\n_atom_site_fract_y\\n_atom_site_fract_z\\nSr a
  0.000 ...)
18. matra_generate({"n": 20, "condition": "EHULL_DISC EH0 STOP ELMS Sr Zr Cu S STOP STOICH 2
  1 2 5", "temperature": 0.7})
19. matra_generate({"n": 16, "condition": "EHULL_DISC EH0 STOP ELMS Ca Zr Cu S STOP STOICH 2
  1 2 5", "temperature": 0.8})
20. reactions_decompose({"composition": "Ca2ZrCu2S5"})
21. alexandria_find_structure({"query": "CaS", "limit": 1})
22. tugaff_relax({"format": "cif", "structures": "[...]"})
23. tugasp_predict({"properties": ["mbj_gap", "band_gap_ind", "e_above_hull"], "structures":
  "[...]"})
\end{Verbatim}
Answer:
\begin{Verbatim}[fontsize=\footnotesize]
I found no generated structure that beats Sr2ZrCu2S5. The one new structure worth keeping is
  a **new SrZrCu2S4 polymorph** ("structure I"). It is a possible backup if Sr2ZrCu2S5's gap
  turns out too small, but it is slightly unstable and not yet confirmed by DFT.

## What I did
- **Generated** 96 candidate structures with Matra in five batches. Only about half came out
  valid, and the model often ignored the composition I asked for. That left 23 usable
  structures: 6 Sr2ZrCu2S5 (of which 3 were distinct), 11 SrZrCu2S4, 6 Ca2ZrCu2S5.
- **Relaxed** them with TugaFF, together with the reference phases needed to say what each
  composition would decompose into.
- **Checked the potential first.** It reproduces DFT closely: Sr2ZrCu2S5 comes out 26.6
  meV/atom below its decomposition products (DFT: 27-28), and the database SrZrCu2S4 113
  meV/atom above the hull (DFT: 104). That agreement is why I trust its ranking.
- **Predicted gaps** with the ML model for the best structures.

## Results
Energies are relative to the hull or to the database structure; the ML-MBJ gap is the
  model's prediction, which runs low.

| Structure | Stability (meV/atom) | ML-MBJ gap (eV) | Verdict |
|---|---|---|---|
| **Sr2ZrCu2S5, database structure** | **-27 (on hull)** | 0.68 | still the best |
| Sr2ZrCu2S5, Matra C2/m | same as database | - | rediscovered the known structure |
| Sr2ZrCu2S5, Matra P2/m, Cmmm, P4/mmm | +137, +160, +464 vs database | - | worse |
| **SrZrCu2S4, new structure I (C2/m)** | **+18** (ML model: +15) | **1.21** | best new
  structure; slightly unstable |
| SrZrCu2S4, new F and G | +76, +80 | 1.68 (F) | too unstable |
| SrZrCu2S4, database I222 | +113 | - | for reference |
| Ca2ZrCu2S5, Sr2ZrCu2S5 structure | +17 | 0.58 | unstable and smaller gap |
| Ca2ZrCu2S5, Matra I4/mmm and Cmmm | +110, +103 | - | worse |

## What structure I adds
- Its predicted gap is about 0.5 eV wider than Sr2ZrCu2S5's on the same model. Applying the
  same corrections as before (the Ba2Cu2ThS5 calibration), my rough estimates are:
  - Sr2ZrCu2S5: about 1.0-1.5 eV
  - SrZrCu2S4-I: about 1.5-2.0 eV

  The two compounds bracket your 1.3-1.6 eV window from either side.
- It is 18 meV/atom above the hull, and its likely decomposition products are Sr2ZrCu2S5 +
  ZrCu2S3. It could only be made as a trapped metastable phase (by quenching or
  low-temperature routes), and 18 meV/atom is within the potential's typical error, so only
  DFT can decide whether it is stable.
- It is about 95 meV/atom below the only SrZrCu2S4 structure in Alexandria, so it is new for
  this composition.

## Recommendation
Keep Sr2ZrCu2S5 as the target. The HSE06 calculation I recommended earlier now does double
  duty:
- If Sr2ZrCu2S5's gap is 1.3 eV or more, go ahead with it as planned.
- If it comes out low, run DFT (PBE relaxation, hull check, HSE06 gap) on SrZrCu2S4-I next.
  In the same synthesis, a slightly Zr-rich charge would show whether that phase forms at
  all.

## Limits
- The sampling was small and the generator was unreliable about stoichiometry. A wider
  search would need far more samples, which is better run as a batch job.
- All energies here come from the ML potential and all gaps from ML models. Nothing in this
  round is DFT.
- I couldn't save files (file writing is disabled), so here is the relaxed structure I in
  full:

```
SrZrCu2S4-I (TugaFF-relaxed, P1 setting, ~C2/m)
a=15.3447 b=3.8399 c=7.4231 A, alpha=90.00 beta=125.41 gamma=90.01
Sr 0.0000 0.5003 0.5008 | Sr 0.5000 0.0004 0.4992
Zr 0.9997 0.9985 0.0006 | Zr 0.5002 0.4991 0.9994
Cu 0.8156 0.5000 0.6499 | Cu 0.1843 0.5025 0.3520
Cu 0.3158 0.0021 0.6480 | Cu 0.6845 0.0001 0.3503
S 0.6256 0.4998 0.4361 | S 0.3745 0.5007 0.5626
S 0.3893 0.9989 0.0156 | S 0.6111 0.9994 0.9829
S 0.1254 0.0008 0.4374 | S 0.8745 0.9998 0.5639
S 0.8888 0.4991 0.0168 | S 0.1107 0.4985 0.9845
```
\end{Verbatim}